\documentclass[journal]{IEEEtran}
\ifCLASSINFOpdf
\else
\fi

\usepackage{bm}
\usepackage{graphicx}
\usepackage{indentfirst}
\usepackage{epsfig}
\usepackage{amsfonts}
\usepackage{amsmath}
\usepackage{url}
\usepackage{color}
 \usepackage{algorithm}
\usepackage{algpseudocode}
\usepackage{cancel}
\usepackage{graphicx}
\usepackage{mathcomp}
\usepackage{footnote}
\usepackage{cite}
\usepackage{mathtools}
\usepackage{amssymb}
\usepackage{amsmath}
\usepackage{amsthm}
\newtheorem{theorem}{Theorem}

\usepackage{mathrsfs}
\usepackage{amsfonts}
\usepackage{graphicx}
\usepackage[justification=centering]{caption}
\usepackage{multirow}
\usepackage{hhline}
\usepackage{bigstrut}
\usepackage{cases}
\usepackage{verbatim}
\usepackage[utf8x]{inputenc}
\usepackage[T1]{fontenc}
\usepackage{xcolor}
\usepackage{epstopdf}

\begin{document}
\title{Joint Investment and Operation of Microgrid}

\author{Hao~Wang,~\IEEEmembership{Student~Member,~IEEE,}
        and~Jianwei~Huang,~\IEEEmembership{Senior~Member,~IEEE}

\thanks{This work is supported by a grant from the Research Grants Council of the Hong Kong Special Administrative Region, China, under Theme-based Research Scheme through Project No. T23-407/13-N. Part of the results have appeared in IEEE SmartGridComm 2014 \cite{ourpaper}.}
\thanks{H. Wang and J. Huang (corresponding author) are with the Network Communications and Economics Lab (NCEL), Department of Information Engineering, The Chinese University of Hong Kong, Shatin, Hong Kong SAR, China, e-mails: \{haowang, jwhuang\}@ie.cuhk.edu.hk.}
}
\maketitle

\begin{abstract}
In this paper, we propose a theoretical framework for the joint optimization of investment and operation of a microgrid, taking the impact of energy storage, renewable energy integration, and demand response into consideration. We first study the renewable energy generations in Hong kong, and identify the potential benefit of mixed deployment of solar and wind energy generations. Then we model the joint investment and operation as a two-period stochastic programming program. In period-1, the microgrid operator makes the optimal investment decisions on the capacities of solar power generation, wind power generation, and energy storage. In period-2, the operator coordinates the power supply and demand in the microgrid to minimize the operating cost. We design a decentralized algorithm for computing the optimal pricing and power consumption in period-2, based on which we solve the optimal investment problem in period-1. We also study the impact of prediction error of renewable energy generation on the portfolio investment using robust optimization framework. Using realistic meteorological data obtained from the Hong Kong observatory, we numerically characterize the optimal portfolio investment decisions, optimal day-ahead pricing and power scheduling, and demonstrate the advantage of using mixed renewable energy and demand response in terms of reducing investment cost.
\end{abstract}

\begin{IEEEkeywords}
Smart grid, microgrid, renewable energy, solar, wind, energy storage, demand response, investment, pricing. 
\end{IEEEkeywords}

\IEEEpeerreviewmaketitle

\section*{Nomenclature}

\subsection*{Acronyms}
\begin{tabular}{l l} 
	P1 &Period-1 problem for investment \\ 
	P2 &Period-2 problem for operation \\
	Pu$_i$ &User $i$'s cost minimizing problem \\
	EP1 &Equivalent problem of P1 \\ 
	RP1 &Period-1 problem for robust optimization\\ 
	RP2 &Period-2 problem for robust optimization
\end{tabular}

\subsection*{Sets}
\begin{tabular}{l l} 
	$\mathcal{N}$ &Set of users  \\
	$\mathcal{H}$ &Investment horizon \\ 
	$\mathcal{T}$ &Operational horizon \\ 
	$\Omega$ &Set of scenarios
\end{tabular}

\subsection*{Indices}
\begin{tabular}{l l} 
	$i$ &Index of users  \\
	$t$ &Index of hours \\ 
	$\omega$ &Index of scenarios
\end{tabular}

\subsection*{Parameters}
\begin{tabular}{l l} 
	$N$ &Number of electricity users \\
    $T$ &Number of hours in the operational horizon\\
    $D$ &Number of days in the investment horizon\\
    $c_{s}$ &Investment cost of solar power\\
    $c_{w}$ &Investment cost of wind power \\
    $c_{e}$ &Investment cost of energy storage\\
	$B$ &Investment budget\\
	$\beta_{o}$ &Coefficient of the operator's cost\\
	$\beta_{i}$ &Coefficient of user $i$'s discomfort cost\\
	$b^t$ &Aggregate inelastic load in time slot $t$\\
	$D_{i}$ &Total elastic load of user $i$ \\
	$\underline{d}_{i}^{t}$ &Minimum power load of user $i$ in time slot $t$\\
	$\overline{d}_{i}^{t}$ &Maximum power load of user $i$ in time slot $t$\\
	$y_{i}^{t}$ &Preferred power load of user $i$ in time slot $t$\\
	$r_{\max}^{\omega,t}$ &Maximum renewable power in $t$ and $\omega$\\
	$\eta_{s}^{\omega,t}$ &Solar power supply per unit capacity in $t$ and $\omega$\\
	$\eta_{w}^{\omega,t}$ &Wind power supply per unit capacity in $t$ and $\omega$\\
	$r_{c}^{\max}$ &Maximum charging amount per unit capacity \\
	$r_{d}^{\max}$ &Maximum discharging amount per unit capacity \\
	$\eta_{c}$ &Conversion efficiency of charging \\
	$\eta_{d}$ &Conversion efficiency of discharging \\
	$SOC_{\min}$ &Minimum state-of-charge \\
	$SOC_{\max}$ &Maximum state-of-charge \\
	$DOD_{\max}$ &Maximum depth-of-discharge\\
	$e_{s,\min}^{\omega,t}$ &Minimum solar power prediction error in $t$ and $\omega$\\
	$e_{s,\max}^{\omega,t}$ &Maximum solar power prediction error in $t$ and $\omega$\\
	$e_{w,\min}^{\omega,t}$ &Minimum wind power prediction error in $t$ and $\omega$\\
	$e_{w,\max}^{\omega,t}$ &Maximum wind power prediction error in $t$ and $\omega$
\end{tabular}

\subsection*{Variables}
\begin{tabular}{l l} 
	$\alpha_{s}$ &Generation capacity of solar power\\
    $\alpha_{w}$ &Generation capacity of wind power \\
    $\alpha_{e}$ &Capacity of energy storage \\
	$x_{i}^{\omega,t}$ &User $i$'s elastic load in $t$ and $\omega$\\ 
	$r^{\omega,t}$ &Renewable power supply in $t$ and $\omega$\\
	$q^{\omega,t}$ &Grid power procurement in $t$ and $\omega$\\
	$SOC^{\omega,t}$ &State-of-charge of battery in $t$ and $\omega$\\ 
	$r_{c}^{\omega,t}$ &Charging amount in $t$ and $\omega$\\ 
	$r_{d}^{\omega,t}$ &Discharging amount in $t$ and $\omega$\\
	$Q^{\omega,t}$ &Aggregate power supply in $t$ and $\omega$\\
	$p^{\omega,t}$ &Day-ahead price in $t$ and $\omega$\\
	$e_{s}^{\omega,t}$ &Solar power prediction error in $t$ and $\omega$\\
	$e_{w}^{\omega,t}$ &Wind power prediction error in $t$ and $\omega$\\
	$\hat{\eta}_{s}^{\omega,t}$ &Actual solar power generation in $t$ and $\omega$\\
	$\hat{\eta}_{w}^{\omega,t}$ &Actual wind power generation in $t$ and $\omega$
\end{tabular}

\section{Introduction}
Aiming at reducing greenhouse gas emissions and enhancing the power grid reliability, many countries are building new power infrastructures known as the smart grid \cite{smartgrid}. The major features of the smart grid include more distributed power generations (especially from renewable energy sources), smart charging/discharging of energy storage, two-way communications between the utility company and consumers for a better demand side management, and decentralized operations of power grid in the form of microgrids \cite{smartgrid}. It's essential to understand the impact of these new features, and how to make optimal economic and technology decisions on the planning and operation of the smart grid.

Recently, there are many studies on power grid planning (\emph{e.g.} \cite{invest1,invest2}), integration of renewable energy and energy storage (\emph{e.g.} \cite{invest3,invest4,invest5,invest6}), and demand response (\emph{e.g.} \cite{response1,response2,response3}). However, the existing literature did not consider these important issues in a holistic fashion. For example, in \cite{invest1,invest2,invest3,invest4,invest5,invest6}, microgrid planning is studied without considering flexible load and microgrid operation. While in \cite{response1,response2,response3,response4}, only microgrid operation is studied under given microgrid facilities. However, all those new features including renewable energy, storage, and demand response affect the optimal planning and operation of the microgrid, and have to be taken into account at various different time scales. In this paper, we will jointly consider the optimal investment and operation of renewable generation, energy storage, and demand response optimization in the smart grid.

In particular, this paper will focus on the mixed investment in renewable energy (both solar energy and wind energy) and energy storage. The optimal mix of solar and wind energy investment will depend on the stochastic nature of these two sources, which is highly location dependent. Hence we will rely on the meteorological data in Hong Kong to validate the practical relevance of our study. Energy storage provides flexibility in terms of coordinating supply and demand in the microgrid. Through smart charging and discharging of the energy storage, the microgrid operator is able to better utilize the renewable energy generation and reduce dependency on the main grid. The key question we want to answer is the following: \emph{What is the optimal investment portfolio?}

In this paper, we develop a theoretical framework that captures the economic impact of renewable energy, storage, and demand response in the smart grid, and derive the optimal investment strategy and optimal demand response scheme based on realistic data. The main contributions of this paper are as follows.

\begin{itemize}
\item \textit{Correlation and scenarios of renewable energy}: Based on the meteorological data acquired from the Hong Kong Observatory, we study the correlation between solar power and wind power at certain locations of Hong Kong, and suggest mixed renewable energy investment.

\item \textit{Framework development}: We develop a theoretical framework that enables us to derive the optimal investment of mixed enewable generation and energy storage, and the optimal operation in a demand-responsive microgrid. The problem is challenging due to the coupling of decisions of investment and operation at different time scales.

\item \textit{Modeling and solution methods}: We formulate the joint investment and operation problem as a two-period stochastic program. We design a distributed algorithm to attain the optimal power scheduling in period-2, and derive a single-level optimization formulation to solve the optimal investment portfolio in period-1.

\item \textit{Impact of uncertainty in renewable energy}: We analyze the impact of the prediction error of renewable energy generation by the worst-case scenario analysis.

\item \textit{Case studies in Hong Kong}: Numerical studies based on realistic meteorological data illustrate the optimal portfolio investment decisions, and demonstrate the benefits of mixed renewable investment and demand response in terms of saving investment.
\end{itemize}

The remainder of this paper is organized as follows. We review the related work in Section II. Then we analyze the renewable power generation of Hong Kong in Section III, and formulate the system model as a two-period stochastic optimization problem in Section IV. We present the detailed models for period-1 and period-2 in Section V and VI, respectively. We propose the solution method in Section VII. In Section VIII, we analyze the impact of the prediction error of renewable energy generations on the energy portfolio investment. Numerical results are presented in Section IX. This paper is concluded in Section X.

\section{Related work}
There are several related recent studies on power grid planning, integration of renewable energy, and demand response. Specifically, studies in \cite{invest1} and \cite{invest2} examined investment strategies on renewable energy generation through empirical (or numerical) approaches, without considering the power scheduling operation. Studies in \cite{invest3} and \cite{invest4} formulated cost minimization problems to determine the optimal investment of solar-storage system and wind-storage system, respectively. Wang \emph{et al.} in \cite{invest5} considered the optimal planning problem for mixed solar-wind energy in microgrids using robust optimization. Yang and Nehorai in \cite{invest6} formulated a cost optimization problem to decide the optimal capacities for renewable energy generation and energy storage, and solved the problem in a distributed fashion. However, none of these studies took the proactive operation (such as demand response) into consideration. Optimal demand response for residential consumers and energy storage have been studied to derive the proper incentive schemes through either game theoretic models \cite{response4},\cite{response1} or optimization models \cite{response2,response3}. The key idea of these studies is to design incentive mechanisms such that cost-aware users schedule their elastic demands as responses to price changes.

The existing literature focused on either renewable energy and energy storage investment at a large time scale (years), or power scheduling and demand response optimization under given energy capacity at a small time scale (hours and days). However, the decisions at these two different time scales are actually tightly coupled. The portfolio investment determines the time-varying power supply availability and power scheduling flexibility, and thus affects operator's power dispatch and users' demand response at a smaller time scale. Meanwhile, demand response can try to match the demand with the time varying renewable energy supply, and hence can maximize the benefit of renewable energy and even reduce unnecessary investment expenditure at a large time scale. Therefore, the investment and operation at various time scales should be jointly optimized.

Recently, joint optimization of investment and operation has been considered in \cite{joint1,joint2,joint3} to study the wind power investment and network expansion at a transmission-level. Different from \cite{joint1,joint2,joint3}, our work aims to study the energy portfolio investment in microgrids at a distribution-level. Specifically, we consider a holistic configuration of the microgrid to incorporate the latest technologies of smart grid, which includes not only renewable energy but also energy storage and demand response. Moreover, we construct an investment portfolio that consists of different technologies (\emph{e.g.}, solar power, wind power, and storage), and provide a systematic framework to jointly determine the optimal portfolio investment strategy and optimal pricing scheme for demand response.

\section{Solar power and wind power in Hong Kong}
\begin{figure}[t]
	\centering
	\includegraphics[width=7cm]{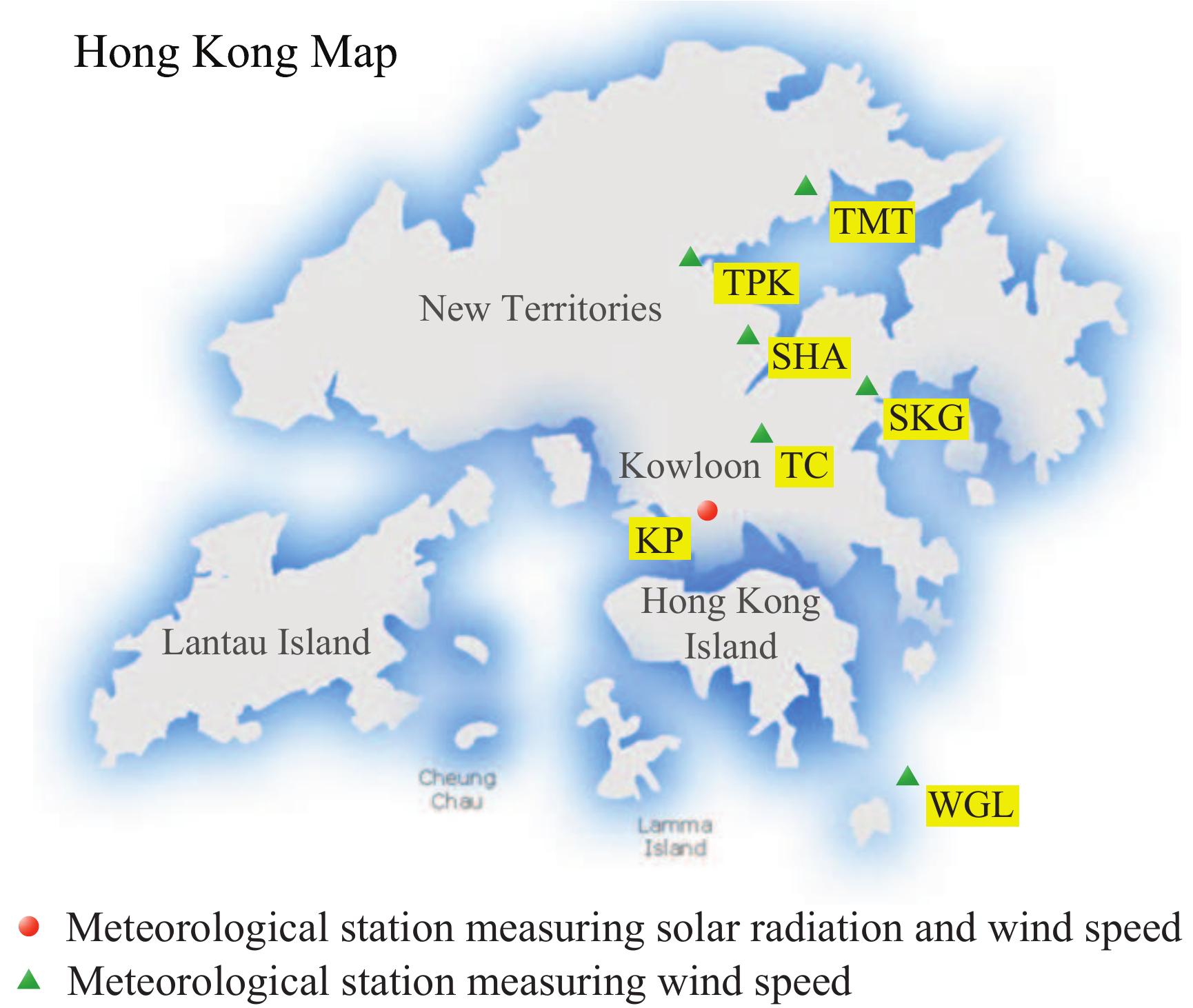}
	\caption{\label{fig-hk} Locations of meteorological stations}
\end{figure}

Aiming at studying the renewable power patterns in Hong Kong, we acquire meteorological data from the Hong Kong Observatory. The data include the hourly solar radiation in King's Park (KP) of Hong Kong, and hourly wind speed at seven different locations (KP, TMT, TPK, SHA, SKG, TC, WGL) of Hong Kong, as shown in Fig. \ref{fig-hk}. Based on the data, we analyze the correlations of solar power and wind power generations across different locations of Hong Kong, which motivate us to study the mixed renewable energy investment in microgrids.

\subsection{Correlation between solar power and wind power}
Both solar power and wind power are intermittent power sources, and their stochastic features could be highly locational-dependent. Since Hong Kong is relatively small, we assume that the solar radiation is the same across Hong Kong and is represented by the data measured in KP. The wind power, however, has clearly different patterns at different locations. 

In this paper, we will focus on one year of meteorological data (from Sep. 1 2012 to Aug. 31 2013) to study the solar power and wind power productions. Based on the solar power model \cite{model1} and wind power model \cite{model2}, we calculate the hourly solar and wind power productions in $365$ days based on the measurement of data of solar radiation and wind speed.\footnote{The technical parameters of the solar power model and wind power model are shown in the online technical report \cite{report}.} To study the statistical correlation between the hourly solar and wind power productions over one year, we calculate the sample correlation coefficient \cite{sweden} as
\begin{align*}
 \rho_{X,Y} = \frac{\sum_{k} \left(X(k)-\bar{X} \right) \left(Y(k)-\bar{Y} \right)} {\sqrt{\sum_{k} \left(X(k)-\bar{X} \right)^{2}}  \sqrt{\sum_{k} \left(Y(k)-\bar{Y} \right)^{2}} },
\end{align*}
where $X$ and $Y$ are data series with $k=1,...,K$ terms, $\bar{X}$ and $\bar{Y}$ are the mean values of $X$ and $Y$, respectively, and $\rho_{X,Y}$ measures the correlation coefficient between $X$ and $Y$. We substitute the one-year hourly solar power production into $X$, and the one-year hourly wind power production of each location into $Y$, and calculate their correlations. We find that the wind powers in four locations (KP, TPK, SHA, SKG) of Hong Kong have positive correlations with solar power, while the correlation is negative in two locations (TC, WGL), and the correlation is close to zero in location TMT. 

Motivated by the Markowitz portfolio selection theory in Finance \cite{Markowitz}, we will study the mixed investment strategy of solar power and wind power in the following two locations: TC and SKG. Specifically, TC and SKG are two representative examples for negative and positive correlations (with correlation coefficient $-0.22$ and $0.15$) between solar and wind power generations, respectively.

\subsection{Scenario generation of solar power and wind power}
To study the mixed investment of solar power and wind power, we need to model the solar power and wind power generations. Usually, the renewable energy investment is made for years of operation. Therefore, we use one-year historical data to empirically model the distributions of solar and wind power generations, similar as \cite{joint1}, and assume that the future renewable generations in each year follow the same distribution of the one-year historical data. Each daily power production realization (solar power production in KP, wind power productions in TC and SKG) is called a \emph{scenario}, and thus we obtain 365 scenarios for solar power and wind power respectively. As a large number of scenarios will reduce the computational tractability of the investment optimization problem, it is useful to choose a smaller subset of scenarios that can well approximate the original entire scenario set. Such technique has been widely used in economics and engineering research \cite{scenario1,scenario2} for the purpose of modeling stochastic processes.
\begin{figure*}[t]
\begin{center}
\begin{minipage}[c]{0.32\textwidth}
  \includegraphics[width=5.2cm]{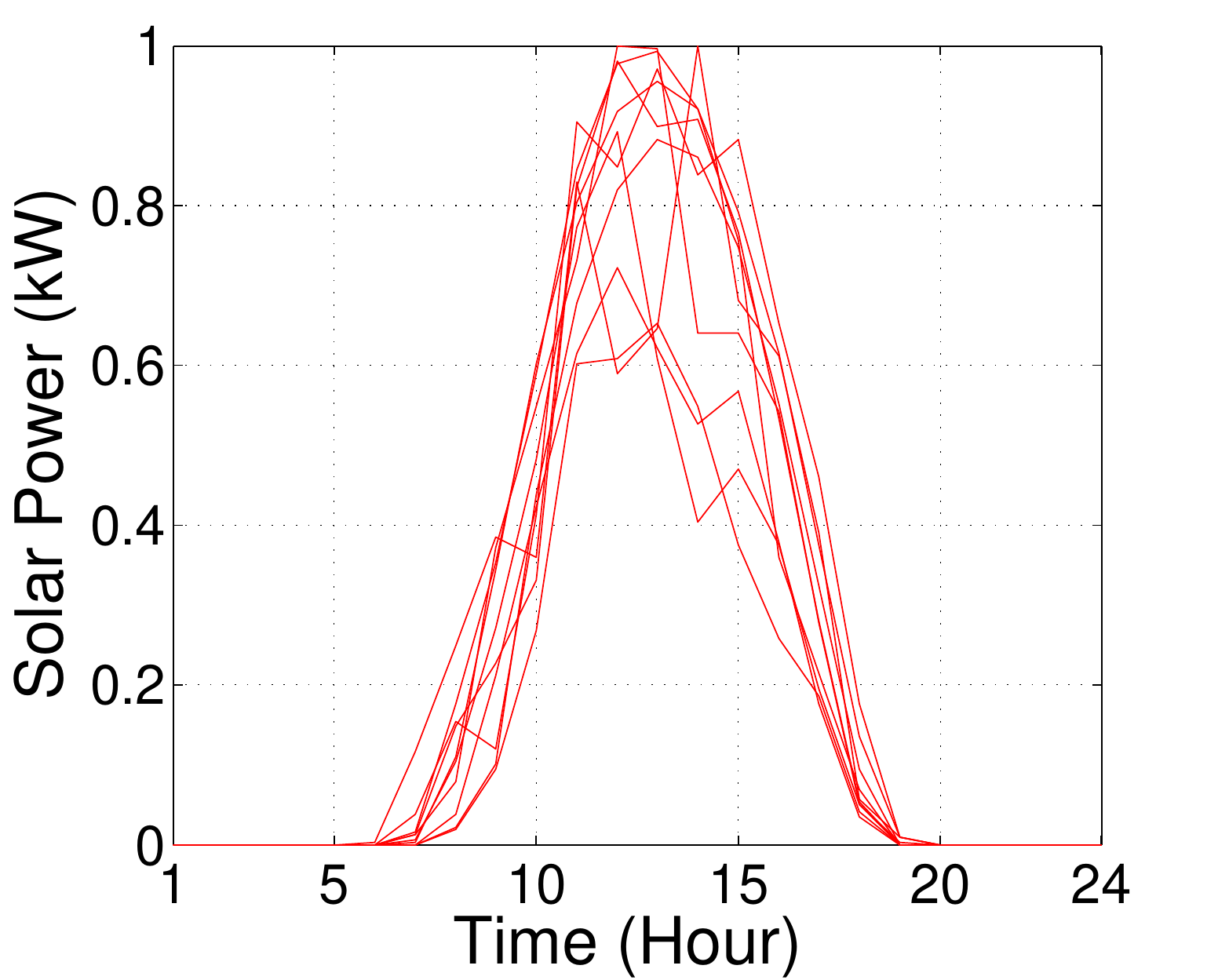}
  \caption{\label{fig-power-KP}Solar power scenarios in KP \\(per 1kW capacity)}
\end{minipage}
\begin{minipage}[c]{0.32\textwidth}
   \includegraphics[width=5.2cm]{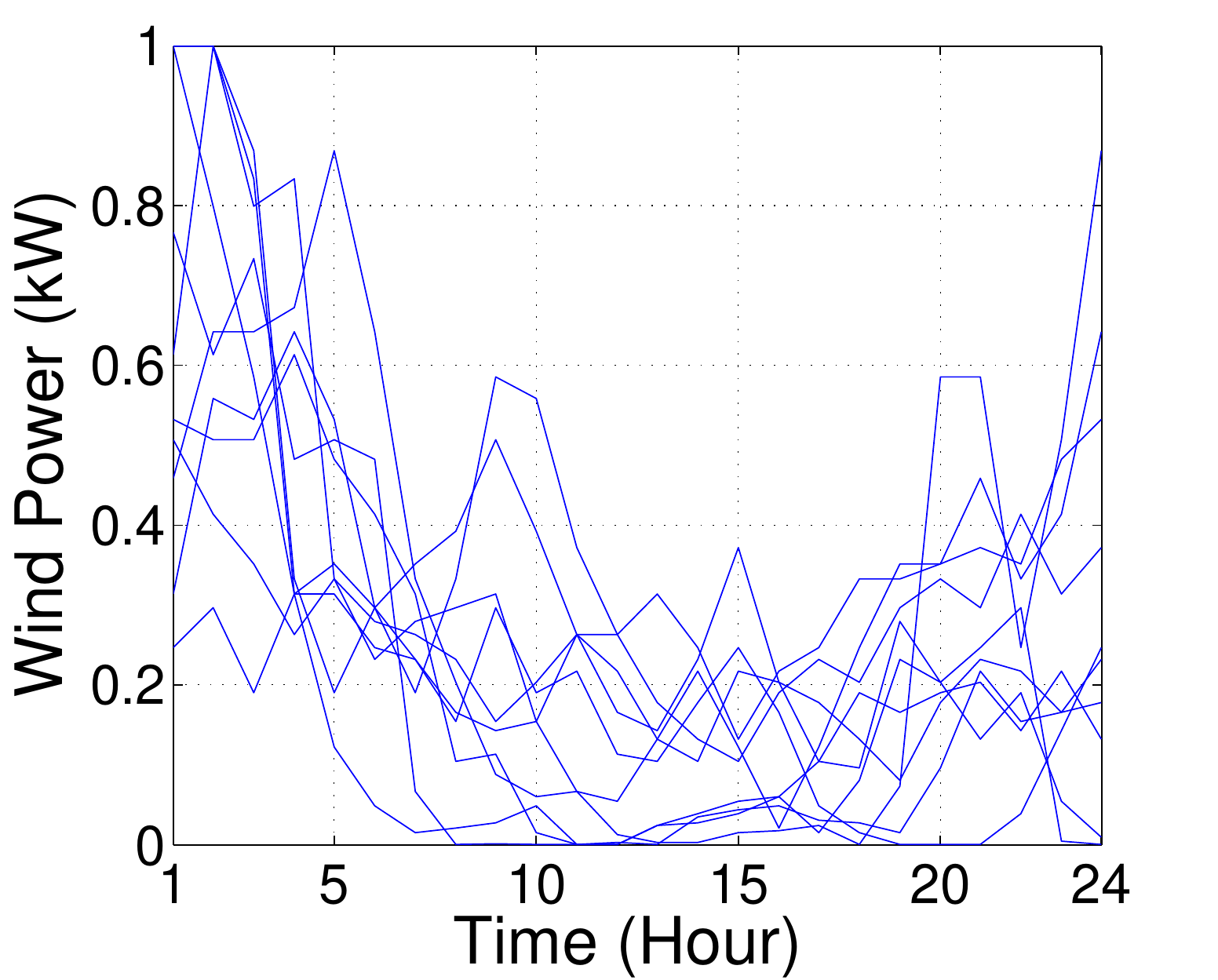}
   \caption{\label{fig-power-TC}Wind power scenarios in TC \\(per 1kW capacity)}  
\end{minipage}
\begin{minipage}[c]{0.32\textwidth}
  \includegraphics[width=5.2cm]{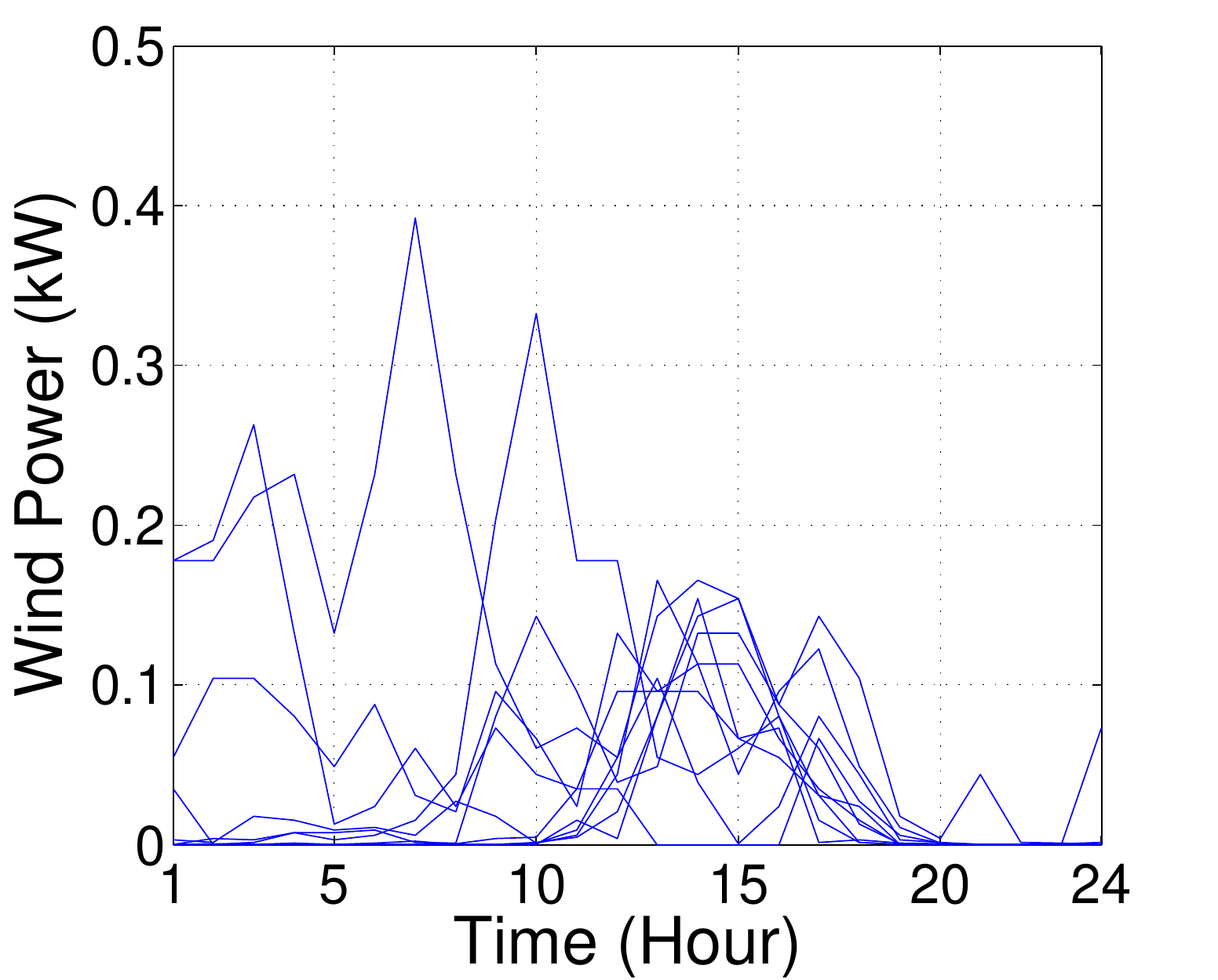}
  \caption{\label{fig-power-SKG}Wind power scenarios in SKG \\(per 1kW capacity)}
\end{minipage}
\end{center}
\end{figure*}

We applied the forward scenario reduction algorithm \cite{scenario2} to find a best scenario subset, and assign new probabilities to the smaller number scenarios. The key idea is to select a subset of scenarios to preserve, such that the corresponding reduced probability measure is the closest to the original measure. We set the number of preserved scenarios as $10$,\footnote{The persevered number of scenarios depends on the tradeoff between performance and computational complexity in practice.} and generate selected scenarios for the solar power production in KP (which we assume is the same as in TC and SKG since Hong Kong is relatively small geographically) and wind power productions in TC and SKG, shown in Fig. \ref{fig-power-KP}, \ref{fig-power-TC}, and \ref{fig-power-SKG}. Therefore, we have a set of $10$ scenarios denoted as $\Omega$ for the renewable energy generation, and each renewable generation scenario $\omega \in \Omega$ consists of solar power and wind power productions in TC and SKG.\footnote{For the detailed scenario generation and reduction, please refer to the technical report \cite{report}.} Comparing Fig. \ref{fig-power-TC} and \ref{fig-power-SKG} with Fig. \ref{fig-power-KP}, we can see that the solar power has a peak at noontime, while wind power productions show dramatic locational differences. Wind power in TC is often adequate during night time, while wind power in SKG reaches a higher output level during day time. Therefore, solar power and wind power have high locational dependence, which motivates us to study the optimal mixed investment of both cases. The data we use can be found at \cite{releasedata}.

\section{System Overview}
\begin{figure}[tbhp]
\centering
  \includegraphics[width=8cm]{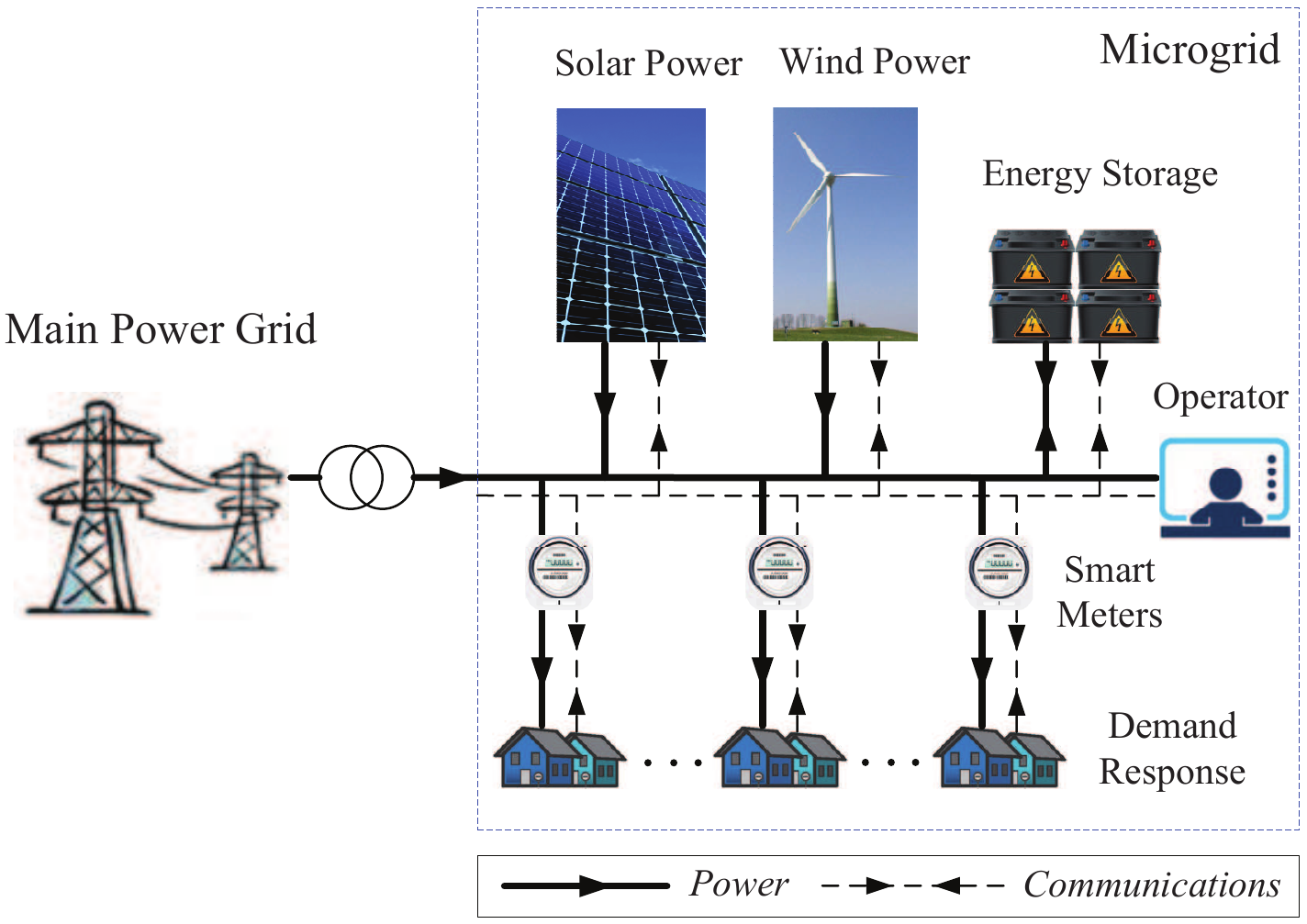}
  \caption{\label{fig_system} System model}
\end{figure}

In this section, we present the system model for the joint investment and operation problem. Fig. \ref{fig_system} illustrates a typical microgrid, which connects to the main power grid, and consists of different local power supplies and responsive demand.

Within the microgrid, there is a local generation system, consisting of solar and wind renewable power generation. An energy storage device is also implemented in the microgrid to charge and discharge power properly to reduce the system operating cost. The demand side consists of a set of electricity users $\mathcal{N}=\{1,...,N\}$, and each user $i \in \mathcal{N}$ is equipped with a smart meter and energy scheduling module. The operator runs the microgrid, determines the investment in renewable energy and energy storage capacities at a large time scale (years), as well as energy prices at a small time scale (hours in one day). The users determine their energy consumptions based on the prices set by the operator (as the demand response).

\begin{figure}[h]
	\centering
	\includegraphics[width=6cm]{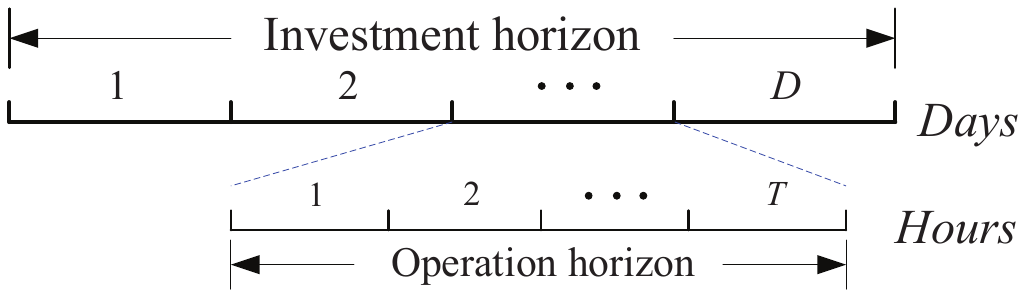}
	\caption{\label{fig-horizon} Investment and operation horizons}
\end{figure}

From the operator's perspective, it needs to decide the optimal capacity investment and power scheduling. Fig. \ref{fig-horizon} depicts the investment and operation horizons. An investment horizon usually corresponds to several years. The operation horizon is one day, which includes $\mathcal{T}=\{1,...,T\}$ of $T$ time slots (say 24 hours). To model both investment and operation, we propose a two-period stochastic program that jointly optimizes capacity investment and power scheduling in the microgrid. Specifically, the period-1 problem is a long-term capacity investment problem, with the objective of minimizing the expected overall cost over an investment horizon $\mathcal{H}=\{1,...,D\}$ of $D$ days, subject to a budget constraint. The period-2 problem is a power scheduling problem, with the objective of minimizing the operating costs of both operator and users under a specific realization of renewable power generation within a smaller time window.
\begin{figure}[h]
	\centering
	\includegraphics[width=6cm]{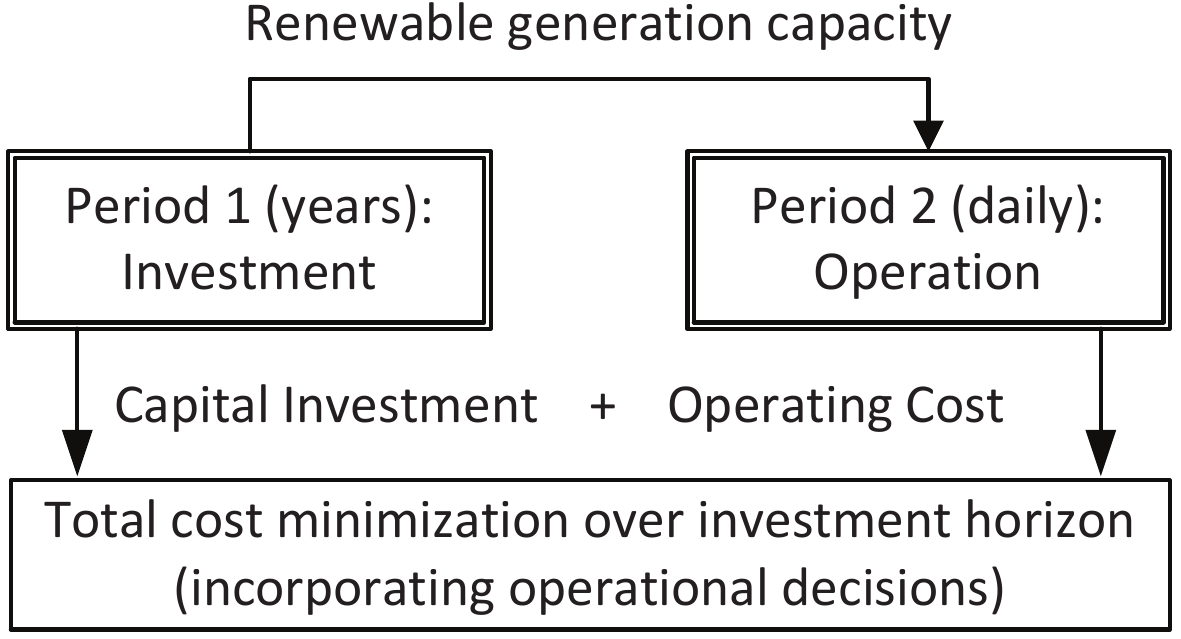}
	\caption{\label{fig-twoperiods} Two-periods optimization problem}
\end{figure}

Note that the investment and operation problems are closely coupled, as the investment decision affects the operational cost, and the expected recurring operational cost affects the investment decision. Fig. \ref{fig-twoperiods} shows the connection between period-1 investment problem and period-2 operational problem. In period-2, the optimal power scheduling is affected by both power supply and demand, and in particular the renewable power supply depends on the capacity investment decision made in period-1. On the other hand, the overall cost in period-1 includes both the one-time investment cost and the recurring operating cost of all $D$ days in period-2. Hence the decisions in two periods are coupled. In the following, we will first formulate the operation problem in period-2, and then formulate the investment problem in period-1.

\section{Period-2 problem for microgrid operations}
In this section, we first present the models of users and the operator, and then formulate the operator's operating cost minimization problem.

\subsection{User's model}
We classify each user's load into two types: the elastic load and the inelastic load. The elastic load corresponds to the energy usage of those appliances such as electric vehicles, washing machines, and HVAC (heating, ventilation and air Conditioning) systems, as a user may shift the flexible load over time. The inelastic load corresponds to the energy usage of appliances such as lighting, refrigerators, and such load cannot be easily shifted over time. The demand response can only control the elastic load. We denote the corresponding decisions as $\boldsymbol{x}_{i}^{\omega} = \{ x_{i}^{\omega,t} ,~\forall t \in \mathcal{T} \}$, where $x_{i}^{\omega,t}$ is user $i$'s elastic energy consumption in time slot $t \in \mathcal{T}$ under renewable energy generation scenario $\omega$.

The elastic load scheduling for all users needs to satisfy the following two constraints:
\begin{align}
& \underline{d}_{i}^{t}\leq x_{i}^{\omega,t}\leq\overline{d}_{i}^{t},\;\forall t\in\mathcal{T},~i \in \mathcal{N}, ~~ \label{loadconstraint1} \\
 & \sum_{t \in \mathcal{T}} x_{i}^{\omega,t} = D_{i},~i \in \mathcal{N}. \label{loadconstraint2}
\end{align}
Constraint (\ref{loadconstraint1}) provides a minimum power consumption $\underline{d}_{i}^{t}$ and a maximum power consumption $\overline{d}_{i}^{t}$ for the user $i$ in each time slot $t$.\footnote{Constraint (\ref{loadconstraint1}) is general and can model the staring time and ending time of an elastic load by setting proper parameters.} Constraint (\ref{loadconstraint2}) corresponds to the total elastic energy demand $D_{i}$ for user $i$ in the entire operation horizon.

We further introduce a discomfort cost $C_{i}(\cdot)$, which measures user $i$'s experience under $\boldsymbol{x}_i^{\omega} = \{ x_i^{\omega,t},~\forall t \in \mathcal{T} \}$ which deviates from his preferred power consumption $\boldsymbol{y}_{i} = \{ y_{i}^{t} ,~\forall t \in \mathcal{T} \}$ under a time-independent flat price environment. We assume that the operator can control the demand to minimize the operating cost. Later in Section VII, we will discuss how such control can be realized through a properly designed pricing mechanism. If the operator sets time-varying prices, energy users will schedule their elastic loads to minimize users' costs, \emph{i.e.} shifting power consumption from high price time slots to low price time slots. The corresponding discomfort (or inconvenience) \cite{discomfort} is
\begin{align}
& C_{i}(\boldsymbol{x}_{i}^{\omega}) = \beta_{i} \sum_{t \in \mathcal{T}} \left(x_{i}^{\omega,t}-y_{i}^{t}\right)^{2},
\end{align}
where $\beta_{i}$ is the coefficient of discomfort cost, which indicates the sensitivity of user $i$ towards its deviation from the preferred power consumption.

\subsection{Operator's model}
We assume that the operator can predict the renewable energy production scenario $\omega$ accurately at the beginning of an operation horizon (a day).\footnote{The short-run (day-ahead) renewable energy forecast can be quite accurate in practice \cite{forecast1,forecast2}. We will also consider the impact of prediction error in Section VIII. In addition, this paper focuses on the microgrid energy management instead of the transient dynamics and stability of the sysetem. Therefore, we assume that the microgrid operator is able to manage the intermittency of renewable generations, such that the stability of the microgrid system can always be guaranteed.} In scenario $\omega$ and each time slot $t$, the operator determines the renewable power supply, conventional power procurement, and energy storage charging and discharging to meet the total users' demand, which consists of the elastic power consumption $\boldsymbol{x}_{i}^{\omega}$ from each user $i \in \mathcal{N}$ and the aggregate inelastic load $\{ b^t,~t \in \mathcal{T} \}$ of all the users. 

\subsubsection{Power supply}
The renewable power supply $\boldsymbol{r}^{\omega} = \{ r^{\omega,t},~\forall t \in \mathcal{T} \}$ and conventional power procurement $\boldsymbol{q}^{\omega} = \{ q^{\omega,t} ,~\forall t \in \mathcal{T} \}$ should satisfy the following constraints:
\begin{align}
 & 0 \leq r^{\omega,t} \leq r_{\max}^{\omega,t} , \forall t\in\mathcal{T}, \label{supplyconstraint1} \\
 & q^{\omega,t} \geq 0 , \forall t\in\mathcal{T}, \label{supplyconstraint2}
 \end{align}
where $r_{\max}^{\omega,t}$ in constraint (\ref{supplyconstraint1}) depends on the invested capacities of solar power $\alpha_{s}$ and wind power $\alpha_{w}$, which are the operator's decision variables in period-1. For each unit of invested solar capacity and wind capacity, the corresponding solar power and wind power in scenario $\omega$ and time $t$ will be $\boldsymbol{\eta}_{s}^{\omega} = \{ \eta_{s}^{\omega,t} ,\forall t \in \mathcal{T} \}$ and $\boldsymbol{\eta}_{w}^{\omega} = \{ \eta_{w}^{\omega,t} ,\forall t \in \mathcal{T} \}$, respectively. Hence we have  $r_{\max}^{\omega,t} = \eta_{s}^{\omega,t} \alpha_{s} + \eta_{w}^{\omega,t} \alpha_{w}$. Constraint (\ref{supplyconstraint2}) means that the operator can only purchase power from the main grid, but cannot sell power to the main grid, assuming that the main grid does not accommodate distributed generations in the microgrid. We assume that the main grid has adequate power to meet the demand of the microgrid, hence there is no upper-bound of $q^{\omega,t}$ in (\ref{supplyconstraint2}). Different from the conventional power generation, the renewable power generation does not consume fuel sources, so we assume zero cost of generating renewable power \cite{invest3}. Therefore, the operator will try to use as much renewable power as possible to meet the demand.

\subsubsection{Energy storage}
It's well-known that energy storage (such as batteries) can smooth out the intermittent renewable power generation, match demand and supply by smart charge/discharge, and exploit time-varying energy generation costs for arbitrage. In our paper, energy storage is regarded as an energy asset in the investment portfolio, as it can complement time-varying renewable energy generation to provide relatively stable power supply, and can also work in parallel with demand response programs to help balance demand and supply. We assume that the microgrid operator has decided in period-1 to install energy storage devices with a total capacity $\alpha_{e}$, and let $SOC^{\omega,t}$, $r_{c}^{\omega,t}$, and $r_{d}^{\omega,t}$ denote the state-of-charge of the storage, charging amount, discharging amount in time slot $t$ and scenario $\omega$, respectively.

First, the energy charging and discharging amounts are bounded, and satisfy the following constraints:
\begin{align}
 & 0 \leq r_{c}^{\omega,t} \leq \alpha_{e} r_{c}^{\max},~\forall t\in\mathcal{T}, \label{constraint-storage1} \\
 & 0 \leq r_{d}^{\omega,t} \leq \alpha_{e} r_{d}^{ \max},~\forall t\in\mathcal{T}, \label{constraint-storage2}  
\end{align}
where $r_{c}^{ \max} >0$ and $r_{d}^{ \max} >0$ denote the maximum charging and discharging amount per unit capacity of the energy storage, respectively. Hence, $\alpha_{e} r_{c}^{\max}$ and $\alpha_{e} r_{d}^{ \max}$ indicate the maximum charging and discharging amount after the operator decides to deploy energy storage facilities with the capacity of $\alpha_{e}$.

Second, there are power losses when electricity is charged into and discharged from the battery. We denote $\eta_{c} \in \left[ 0,1 \right]$ and $\eta_{d} \in \left[ 0,1 \right]$ as the conversion efficiencies of charging and discharging. Therefore, we obtain the energy storage dynamics of the microgrid $i$ in time slot $t$ as
\begin{align}
& SOC^{\omega,t} = SOC^{\omega,t-1} + \frac{\eta_{c} r_{c}^{\omega,t}}{\alpha_{e}} - \frac{r_{d}^{\omega,t}}{\eta_{d} \alpha_{e}} ~\forall t \in \mathcal{T}, \label{constraint-storage3} \\
& SOC_{\min} \leq SOC^{\omega,t} \leq SOC_{\max}, ~\forall t \in \mathcal{T}, \label{constraint-storage4} \\
& SOC^{\omega,0} = SOC^{\omega,T}, \label{constraint-storage5}
\end{align}
where $SOC^{\omega,t}$ evolves with charging and discharging of the battery according to (\ref{constraint-storage3}). It is shown in (\ref{constraint-storage4}) that $SOC^{\omega,t}$ is bounded between $SOC_{\min}$ and $SOC_{\max}$, which are lower and upper bounds \cite{SOC} for the level of energy storage in percentage, respectively. For example, we can set $SOC_{\max}$ as $100 \%$, which means the battery can be charged to reach its full capacity. We can set $SOC_{\min} = 1 - DOD_{\max} $, where $DOD_{\max}$ is the maximum depth-of-discharge (DOD) allowed.\footnote{DOD is defined as the ratio of maximum discharge to the battery capacity \cite{DOD}. Usually, the lifetime of a battery can be measured by the number of charge-discharge cycles it can sustain at a given $DoD_{\max}$. High $DOD_{\max}$ causes fast depreciation of battery storage. Therefore, we set a low $DOD_{\max}$ (thus a high $SOC_{\min}$) for the battery operation to reduce the impact of battery degradation, so as to make sure that the lifetime of battery is no shorter than the investment horizon.} Moreover, we set the terminal state-of-charge $SOC^{\omega,T}$ at the end of each day to be equal to its initial value $SOC^{\omega,0}$ at the beginning of each day, such that the battery can be operated independently across days.

\subsubsection{Operator's cost}
The power supply and demand should satisfy the following power balance constraint in time slot $t$.
\begin{align}
 & r^{\omega,t} + q^{\omega,t} + r_{d}^{\omega,t} = r_{c}^{\omega,t} + b^{t} + \sum_{i \in \mathcal{N}} x_{i}^{\omega,t},\;\forall t\in\mathcal{T}, \label{supplyconstraint3}
\end{align}

We let $\boldsymbol{Q}^{\omega}=\{ Q^{\omega,t},~ t \in \mathcal{T} \} $ denote the aggregate supply, \emph{i.e.}, 
\begin{align}
 & Q^{\omega,t} = r^{\omega,t} + q^{\omega,t} \geq 0,\;\forall t\in\mathcal{T}, \label{supplyconstraint4}
\end{align}
and we can rewrite the power balance constraint (\ref{supplyconstraint3}) as follows:
\begin{align}
 & Q^{\omega,t} = b^{t} + \sum_{i \in \mathcal{N}} x_{i}^{\omega,t} + r_{c}^{\omega,t} - r_{d}^{\omega,t},\;\forall t\in\mathcal{T}. \label{balance}
\end{align}

If the operator has enough renewable generation to meet the aggregate demand, \emph{i.e.} $r_{\max}^{\omega,t} \geq  Q^{\omega,t}$, then there is no need to purchase any conventional power, \emph{i.e.} $q^{\omega,t}=0$. On the other hand, if $r_{\max}^{\omega,t} <  Q^{\omega,t}$, the operator will first use all the renewable power $r^{\omega,t} = r_{\max}^{\omega,t}$, and then purchase conventional power $q^{\omega,t}=Q^{\omega,t}- r_{\max}^{\omega,t} $ to meet the power deficit. The production cost of conventional power has a quadratic form\cite{econ}, and thus we define the operator's cost as
\begin{align}
 & C_{o}(\boldsymbol{Q}^{\omega}) = \beta_{o} \sum_{t \in \mathcal{T}}
 \left[ \left( Q^{\omega,t} - \eta_{s}^{\omega,t}\alpha_{s}-\eta_{w}^{\omega,t}\alpha_{w}
 \right)^{+} \right]^{2}, \label{costo2}
\end{align}
where $(z)^{+}=\max \{z,0 \}$ for any value $z$, and $\beta_{o}$ is the coefficient of the operator's cost.

\subsection{The period-2 problem}
Next we state the period-2 problem, where the operator coordinates aggregate power supply $\boldsymbol{Q}^{\omega}$ and schedules users' power consumptions $\boldsymbol{x}_{i}^{\omega}$ to minimize the operating cost, which consists of the operator's cost $C_{o}(\boldsymbol{Q}^{\omega})$ and all the users' costs $C_{i}(\boldsymbol{x}_{i}^{\omega})$ as follows.
\begin{align*}
& \leftline{\textbf{P2: Operating cost minimization in period-2}}
\end{align*}
\begin{equation*}
\begin{aligned}
& \underset{ \boldsymbol{Q}^{\omega},\boldsymbol{x}_{i}^{\omega} } {\min}
& & C_{o}(\boldsymbol{Q}^{\omega}) + \sum_{i \in \mathcal{N}} C_{i} (\boldsymbol{x}_{i}^{\omega}) \\
& \text{subject to}
& & \text{Constraints \eqref{loadconstraint1}, \eqref{loadconstraint2}, \eqref{constraint-storage1}-\eqref{constraint-storage5}, \eqref{balance}}.
\end{aligned}
\end{equation*}

Problem \textbf{P2} is convex, and can be solved efficiently if a centralized optimization is possible. However, this may not be feasible in practice, as the operator cannot directly control users' power consumptions $\{ \boldsymbol{x}_{i}^{\omega}, \forall i \in \mathcal{N} \}$. We will discuss the design of a pricing scheme to derive the optimal power consumptions of users and the implementation of a decentralized algorithm in Section VII.

\section{Period-1 problem for portfolio investment}
In the period-1 investment problem, the operator needs to determine the capacities of the solar power, wind power, and energy storage facilities ($\alpha_{s}$, $\alpha_{w}$, and $\alpha_{e}$) for the entire investment horizon, subject to a budget constraint $B$.\footnote{we assume that the microgrid operator makes direct investment for the next long period of operation.} These capacity decisions will determine the renewable power production and energy storage flexibility in each day of period-2, and consequently affect the power scheduling and operating cost. The operator wants to make optimal investment decisions to minimize the overall cost, including both the capital investment and the expected operating cost in period-2.

The capital investment cost can be represented as
\begin{align}
& C_{I}(\alpha_{s},\alpha_{w},\alpha_{e}) = c_{s}\alpha_{s}+c_{w}\alpha_{w} +c_{e}\alpha_{e}, \label{costi}
\end{align}
where $c_{s}$, $c_{w}$ and $c_{e}$ denote the investment costs of solar power, wind power and energy storage per $kW$, respectively. The investment cost covers all expenditures, \emph{e.g.}, deployment, installation and maintenance of photovoltaic panel for solar energy, turbine for wind energy, inverters, controllers, and cables.

The daily expected operating cost $\mathbb{E}_{\omega} [ f(\cdot) ]$ is a function of the invested capacities $\alpha_{s}$, $\alpha_{w}$ and $\alpha_{e}$,
\begin{align}
& \mathbb{E}_{\omega \in \Omega}\left[f(\alpha_{s},\alpha_{w},\alpha_{e},\omega)\right] = \sum_{\omega \in \Omega}\pi_{\omega}  f(\alpha_{s},\alpha_{w},\alpha_{e},\omega) ,\label{realized_prob}
\end{align}
where $\omega \in \Omega$ denotes the renewable power scenario with a realization probability $\pi_{\omega}$, which is obtained by the scenario reduction algorithm in Section II. Specifically, the operating cost function in scenario $\omega$ is the optimized objective value (\emph{i.e.} minimized operating cost) of the period-2 problem in scenario $\omega$:
\begin{align}
& f(\alpha_{s},\alpha_{w},\alpha_{e},\omega)=\underset{\boldsymbol{Q}^{\omega}, \boldsymbol{x}_{i}^{\omega}}{\min} 
  [  C_{o}(\boldsymbol{Q}^{\omega}) + \sum_{i \in \mathcal{N}} C_{i} (\boldsymbol{x}_{i}^{\omega}) ], \label{costo3}
\end{align}
where the cost depends on the renewable power supply in scenario $\omega$ and users' demand responses in period-2.

The period-1 optimization problem is subject to a budget constraint and capacity constraints as follows:
\begin{align}
&  c_{s}\alpha_{s}+c_{w}\alpha_{w}+c_{e}\alpha_{e} \leq B, \label{investconstraint1}\\
& \alpha_{s}\geq 0, ~ \alpha_{w}\geq 0, ~\alpha_{e} \geq 0, \label{investconstraint2}
\end{align}
where \eqref{investconstraint1} indicates that the total investment expense cannot be greater than the budget $B$, and the capacity investment must be non-negative.

To summarize, the period-1 problem is as follows.
\begin{align*}
& \leftline{\textbf{P1: Joint investment and operation in period-1}}
\end{align*}
\begin{equation*}
\begin{aligned}
& \underset{\alpha_{s},\alpha_{w},\alpha_{e}}{\min}
&&  C_{I}(\alpha_{s},\alpha_{w},\alpha_{e}) +  D \cdot \mathbb{E}_{\omega \in \Omega}\left[f(\alpha_{s},\alpha_{w},\alpha_{e},\omega)\right] \\
& \text{subject to} 
&& \text{Constraints \eqref{investconstraint1} and \eqref{investconstraint2}},
\end{aligned}
\end{equation*}	
where the objective function consists of capital investment cost $C_{I}$ and expected operating cost $D \cdot \mathbb{E}_{\omega} [ f(\cdot) ]$ under all scenarios $\omega \in \Omega$ over a total $D$ days of operation in the entire investment horizon.

\section{Solution Method}
To solve the above two-period stochastic programming problem, we start with solving the period-2 problem \textbf{P2} using a distributed algorithm. Then we solve the period-1 problem \textbf{P1} to obtain the optimal portfolio investment.

\subsection{Period-2: Optimal power scheduling}
\begin{figure}[tbhp]
\centering
  \includegraphics[width=7cm]{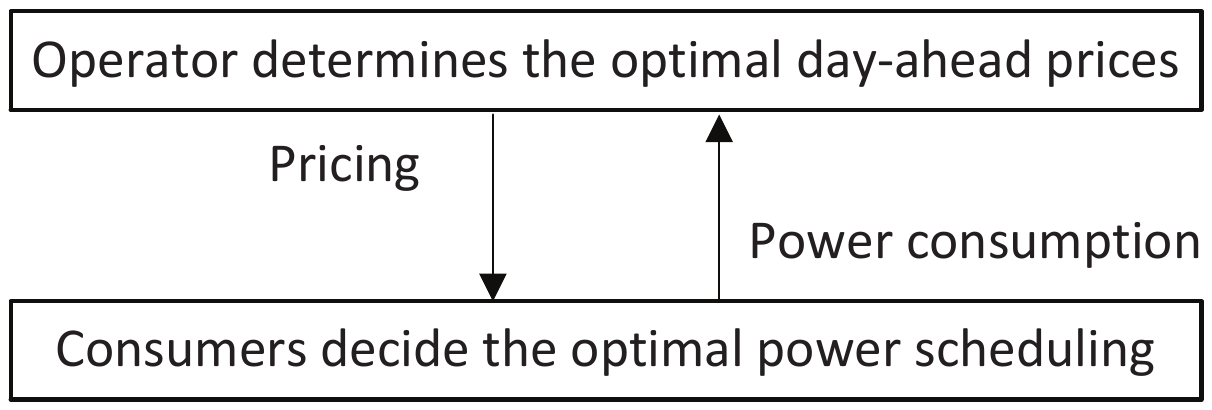}
  \caption{\label{fig-algorithm} Iterative pricing and scheduling}
\end{figure}

As mentioned in Section V, it is not practical for the operator to solve Problem \textbf{P2} centrally and control users' power consumptions directly. Instead, the operator and users compute the price and power consumption in an iterative fashion, as shown in Fig. \ref{fig-algorithm}. 

In particular, the operator can set day-ahead prices $ \boldsymbol{p}^{\omega} = \{ p^{\omega,t},~t \in \mathcal{T} \}$ for the users in scenario $\omega$, and let users choose the proper power scheduling accordingly to minimize their own costs. In the following, we first present a user's total cost minimization problem given the prices. Then we discuss the operator's optimal choices of prices so that the users' power scheduling decisions coincide with the optimal solution of Problem \textbf{P2}.

\subsubsection{User's problem}
In scenario $\omega$, user $i$ receives the price signals $\boldsymbol{p}^{\omega}$ and schedules the power consumption $\boldsymbol{x}_{i}^{\omega}$ to minimize the total cost. Specifically, user $i$'s total cost consists of two parts: energy cost $C_{e}$ and discomfort cost $C_{i}$. The energy cost\footnote{The user cannot change the energy cost related to inelastic load, hence we do not consider that in user's optimization problem.} of user $i$ depends on the price and user $i$'s power consumption, which can be represented as $C_{e}(\boldsymbol{x}_{i}^{\omega}) = \sum_{t \in \mathcal{T}} p^{\omega,t} x_{i}^{\omega,t}$. Therefore, we have the following total cost minimization problem for user $i$ in scenario $\omega$.
\begin{align*}
& \leftline{\textbf{Pu$_i$: User $i$'s cost minimization problem}}
\end{align*}
\begin{equation*}
 \begin{aligned}
 & \underset{\boldsymbol{x}_{i}^{\omega}}{\min}
 && C_{e}(\boldsymbol{x}_{i}^{\omega})
 + C_{i}(\boldsymbol{x}_{i}^{\omega}) \\
 & \text{subject to} 
 && \text{Constraints \eqref{loadconstraint1} and \eqref{loadconstraint2}}, 
 \end{aligned}
\end{equation*}	
where each user solves its optimal power consumption as a response to the price signal set by the microgrid operator.

\subsubsection{Optimal pricing and decentralized algorithm}
We denote $\boldsymbol{p}^{\omega*} = \{ p^{\omega,t*},\forall t \in \mathcal{T} \}$ as the optimal pricing that induces the socially optimal power consumption $\{ \boldsymbol{x}_{i}^{\omega*}, \forall i \in \mathcal{N} \}$ in scenario $\omega$ (\emph{i.e.} the optimal solution of Problem \textbf{P2}). We have the following theorem.
\begin{theorem}
In each scenario $\omega$, the optimal pricing scheme $\boldsymbol{p}^{\omega*}$ that induces the socially optimal power consumptions $\boldsymbol{x}_{i}^{\omega*}$ for each user $i$ satisfies the following relationship,
\end{theorem}
\vspace{-5mm}
\begin{equation}\label{pricing}
p^{\omega,t*}=\begin{cases}
\frac{\partial C_{o}(\boldsymbol{Q}^{\omega})} {\partial Q^{\omega,t}} \Big|_{Q^{\omega,t}=Q^{\omega,t \ast}}, & \text{when}~Q^{\omega,t \ast} > r_{\max}^{\omega,t},\\
0, & \text{when}~Q^{\omega,t \ast} \leq r_{\max}^{\omega,t}.
\end{cases}
\end{equation}

Theorem 1 motivates us to design a decentralized algorithm in \textbf{Algorithm 1}, where the operator sets the day-ahead prices and users respond to the prices by determining their power consumptions. At the beginning of each day, the operator and each user's smart meter compute the hourly electricity prices and the corresponding hourly power consumptions iteratively for the whole-day operation. In \textbf{Algorithm 1}, we consider a sequence of diminishing stepsizes, $\gamma_{k}$'s, which satisfy the following conditions: $\lim_{k\to\infty} \gamma (k)=0$ and $\lim_{k\to\infty} \sum_{k} \gamma (k)=\infty$.\footnote{Problem \textbf{P2} is convex, and the corresponding decentralized algorithm that is executed once a day can coverage fast in the microgrid context with thousands of electricity users.}

\begin{algorithm}[h]
  \caption{Decentralized algorithm in the microgrid}
  \label{alg2}
  \begin{algorithmic}[1]
  \State \textbf{Initialization}: iteration index $k=0$, error tolerance $\epsilon > 0$, stepsize $\gamma(k) > 0$, predicted scenario $\omega$, and $x_{i}^{\omega,t}(0) =y_{i}^{t}$.

\Repeat

  \State \textbf{Operator}: At the $k$-th iteration, the operator collects each user's consumption and computes the aggregate supply $\boldsymbol{Q}^{\omega}$, and sets the prices $p^{\omega,t}(k)$ according to (\ref{pricing}).

  \State \textbf{Users}: Each user $i$ solves Problem \textbf{Pu}$_i$ by updating the power consumption $x_{i}^{\omega,t}(k+1)$ based on the price $p^{\omega,t}(k)$:
 \begin{align*}
  & \hat{x}_{i}^{\omega,t}(k+1)= x_{i}^{\omega,t}(k) 
   - \gamma (k) \left( \frac{\partial C_{i}(\boldsymbol{x_{i}}^{\omega}(k))}{\partial x_{i}^{\omega,t}(k)} + p^{\omega,t}(k) \right).
 \end{align*}
 \State Project the power consumption on the feasible set by solving the following problem:
\begin{equation*}
  \begin{aligned}
  & \underset{\boldsymbol{x}_{i}^{\omega}(k+1)} {\min} 
  && \parallel \boldsymbol{x}_{i}^{\omega}(k+1) - \boldsymbol{\hat{x}}_{i}^{\omega}(k+1) \parallel \\
  & \text{subject to}
  && \text{Constraints \eqref{loadconstraint1} and \eqref{loadconstraint2}}.
  \end{aligned}
\end{equation*}  
  \State $k=k+1$;
  \Until $\parallel \boldsymbol{p}^{\omega}(k) - \boldsymbol{p}^{\omega}(k-1) \parallel \leq \epsilon$.

  \State \textbf{end}
  \end{algorithmic}
\end{algorithm}

We can see that the decentralized \textbf{Algorithm 1} requires minimum information exchange between the operator and users. The users only report their power consumptions to the operator, and the operator broadcasts the prices to all users based on the aggregate power load. There is no need for the users to directly coordinate with each other or to reveal their private information (such as cost function and consumption constraints). The prices set by the operator are the same for all users, and reflect the total power load without disclosing individual user's power consumption. Induced by the optimal pricing scheme set by the microgrid operator, the optimal power consumption of each individual user is the socially optimal power consumption, which minimizes the social cost.
\begin{theorem}
\textbf{Algorithm 1} is a sub-gradient projection algorithm for solving Problem \textbf{P2}, and (with a diminishing stepsize) it converges to the socially optimal price and power consumption $\{ \boldsymbol{p}^{\omega*}, \boldsymbol{x}^{\omega*} \}$ for each $\omega$. 
\end{theorem}

\subsection{Period-1: Optimal energy portfolio investment}
After solving the period-2 problem \textbf{P2}, we solve the period-1 problem \textbf{P1} as follows.
\begin{theorem}
The operator's investment problem \textbf{P1} is equivalent to the following optimization problem \textbf{EP1}:
\vspace{-6mm} 
\end{theorem}
\begin{equation*}
 \begin{aligned}
 & \min
 && \left( c_{s}\alpha_{s}+c_{w}\alpha_{w}+c_{e}\alpha_{e} \right) \\
 && &  + D \cdot \mathbb{E}_{\omega}
 \Big[  C_{o}(\boldsymbol{Q}^{\omega}) + \sum_{i \in \mathcal{N}} C_{i} (\boldsymbol{x}_{i}^{\omega})
 \Big] \\
 & \text{subject to}
 && \text{Constraints \eqref{loadconstraint1}, \eqref{loadconstraint2}, \eqref{constraint-storage1}-\eqref{constraint-storage5}, \eqref{balance}, \eqref{investconstraint1}, \eqref{investconstraint2}}, \\
 & \text{Varaibles:}
 && \alpha_{s},\alpha_{w},\alpha_{e},\boldsymbol{Q}^{\omega},\boldsymbol{x}^{\omega},
 \end{aligned}
\end{equation*} 
in which we assume that the operator can estimate the information about user's power consumption behaviors through a survey or daily operations, and the parameters in (\ref{loadconstraint1}) and (\ref{loadconstraint2}) are known to the operator. Then the operator solves the equivalent problem \textbf{EP1} for the optimal investment in a centralized manner. Note that \textbf{EP1} is a convex quadratic program and can be solved efficiently using a standard interior-point method \cite{convex}.

For the proofs of Theorems, please refer to the online technical report \cite{report}.

\section{Uncertainty in Renewable Generation}

In the previous sections, we have assumed that the operator can predict the solar power generation and wind power generation accurately. Nevertheless, the actual renewable energy generation may deviate from the predicted values due to uncertain weather conditions and different prediction methodologies. Therefore, it is necessary to study the impact of such prediction errors on the optimal operation and investment decisions. In this section, we use the worst-case uncertainty approach \cite{worst-case} to analyze such an issue, by assuming that the uncertain variables are assumed to be bounded in a given uncertainty set. Our aim is to maximize the performance by considering the worst case in the uncertainty set. Note that the robust optimization problem is formulated in period-2 to incorporate the impact of prediction errors of renewable power generations. The period-1 problem, on the other hand, minimizes the overall cost of investment and operation. The detailed formulation is presented in the following.

We define the prediction errors for solar power generation and wind power generation as $e_{s}^{\omega,t}$ and $e_{w}^{\omega,t}$ in time slot $t$ and each scenario $\omega$. Then we can represent the actual solar power generation $\hat{\eta}_{s}^{\omega,t}$ and wind power generation $\hat{\eta}_{w}^{\omega,t}$ as the summation of predicted generations and corresponding errors:
\begin{align*}
 & \hat{\eta}_{s}^{\omega,t} = \eta_{s}^{\omega,t} + e_{s}^{\omega,t},  \\
 & \hat{\eta}_{w}^{\omega,t} = \eta_{w}^{\omega,t} + e_{w}^{\omega,t}. 
\end{align*}

We assume that the prediction errors are bounded in the following uncertainty sets in scenario $\omega$:
\begin{align}
 & \mathcal{E}_{s}^{\omega} = \left\{ e_{s,\min}^{\omega,t} \leq e_{s}^{\omega,t} \leq e_{s, \max}^{\omega,t},~t \in \mathcal{T} \right\}, \label{solarerror} \\
 & \mathcal{E}_{w}^{\omega} = \left\{ e_{w,\min}^{\omega,t} \leq e_{w}^{\omega,t} \leq e_{w,\max}^{\omega,t},~t \in \mathcal{T} \right\}, \label{winderror}
\end{align}
where $e_{s,\min}^{\omega,t}$ and $e_{s, \max}^{\omega,t}$ denote the lower bound and upper bound of the solar power prediction error in time slot $t$ and scenario $\omega$, respectively; $e_{w,\min}^{\omega,t}$ and $e_{w, \max}^{\omega,t}$ denote the lower bound and upper bound of the wind power prediction error in time slot $t$ and scenario $\omega$, respectively.

Firstly, we let $\boldsymbol{e}_{s}^{\omega} = \{ e_{s}^{\omega,t},~ t \in \mathcal{T} \}$ and $\boldsymbol{e}_{w}^{\omega} = \{ e_{w}^{\omega,t},~ t \in \mathcal{T} \}$ denote the prediction error vectors in scenario $\omega$ for the solar power generation and wind power generation, respectively. We define the actual operating cost of the microgrid operator in period-2 as
\begin{align}
 \hat{C}_{o}(\boldsymbol{Q}^{\omega},\boldsymbol{e}_{s}^{\omega},\boldsymbol{e}_{w}^{\omega}) = \beta_{o} \sum_{t \in \mathcal{T}}
 \left[ \left( Q^{\omega,t} - \hat{\eta}_{s}^{\omega,t}\alpha_{s} - \hat{\eta}_{w}^{\omega,t}\alpha_{w}
 \right)^{+} \right]^{2}. \label{actualcosto}
\end{align}

Then, we aim to maximize the worst-case performance of the microgird operation, under all possible prediction errors $\mathcal{E}_{s}^{\omega}$ and $\mathcal{E}_{w}^{\omega}$, using robust optimization techniques \cite{worst-case}. Since the investment cost does not explicitly contain prediction errors, we will focus on the worst operating cost minimizing problem. Based on (\ref{actualcosto}), we formulate the operating cost minimization problem in period-2 as a robust optimization problem denoted as \textbf{RP2}:
\newpage
\begin{align*}
& \leftline{\textbf{RP2: Robust optimization for operating cost minimization}}
\end{align*}
\begin{equation*}
 \begin{aligned}
 & \underset{\boldsymbol{Q}^{\omega},\boldsymbol{x}_{i}^{\omega}}{\min}~
   \underset{\boldsymbol{e}_{s}^{\omega},\boldsymbol{e}_{w}^{\omega}}{\max}
 && \left[  \hat{C}_{o}(\boldsymbol{Q}^{\omega},\boldsymbol{e}_{s}^{\omega},\boldsymbol{e}_{w}^{\omega}) 
 + \sum_{i \in \mathcal{N}} C_{i} (\boldsymbol{x}_{i}^{\omega})
 \right]  \\
 & \text{subject to} 
 && \text{Constraints \eqref{loadconstraint1}, \eqref{loadconstraint2}, \eqref{constraint-storage1}-\eqref{constraint-storage5}, \eqref{balance}, \eqref{solarerror}, \eqref{winderror}}.
 \end{aligned}
\end{equation*}

To solve \textbf{RP2}, we first solve the inner maximization problem of \textbf{RP2}, which aims to maximize the actual operating cost with respect to prediction errors $\boldsymbol{e}_{s}^{\omega}$ and $\boldsymbol{e}_{w}^{\omega}$. Observing the objective function of \textbf{RP2}, we see that users' total cost $\sum_{i \in \mathcal{N}} C_{i} (\boldsymbol{x}_{i}^{\omega})$ is independent of prediction errors. Therefore, we focus on the first term which is the operator's cost. Given the operator's actual power scheduling $\boldsymbol{Q}^{\omega}$, we denote the worst-case prediction errors as
\begin{align}
& \{ \boldsymbol{e}_{s}^{\omega,\ast},~\boldsymbol{e}_{w}^{\omega,\ast} \} = \arg \max_{\boldsymbol{e}_{s}^{\omega} \in \mathcal{E}_{s}^{\omega},~ \boldsymbol{e}_{w}^{\omega} \in \mathcal{E}_{w}^{\omega}}
\hat{C}_{o}(\boldsymbol{Q}^{\omega},\boldsymbol{e}_{s}^{\omega},\boldsymbol{e}_{w}^{\omega}). \label{worstcaseerror}
\end{align}

Since the operator's actual cost $\hat{C}_{o}(\boldsymbol{Q}^{\omega},\boldsymbol{e}_{s}^{\omega},\boldsymbol{e}_{w}^{\omega})$ in (\ref{actualcosto}) is a convex function of $\boldsymbol{e}_{s}^{\omega}$ and $\boldsymbol{e}_{w}^{\omega}$ (under a given $\boldsymbol{Q}^{\omega}$), the optimal solution must hit the boundary of the uncertainty sets $\mathcal{E}_{s}^{\omega}$ and $\mathcal{E}_{w}^{\omega}$. Moreover, we observe that the actual cost of the operator $\hat{C}_{o}(\boldsymbol{Q}^{\omega},\boldsymbol{e}_{s}^{\omega},\boldsymbol{e}_{w}^{\omega})$ is a non-increasing function with respect to $\boldsymbol{e}_{s}^{\omega}$ and $\boldsymbol{e}_{w}^{\omega}$. Thus we have the following theorem:
\begin{theorem}
	The optimal solutions of (\ref{worstcaseerror}) hit the lower bounds, \emph{i.e.} $\boldsymbol{e}_{s}^{\omega,\ast} = \{ e_{s,\min}^{\omega,t},~\forall t \in \mathcal{T} \}$ and $\boldsymbol{e}_{w}^{\omega,\ast} = \{ e_{w,\min}^{\omega,t},~\forall t \in \mathcal{T} \}$.
\end{theorem}

We substitute the worst-case prediction errors $\{ \boldsymbol{e}_{s}^{\omega,\ast},~\boldsymbol{e}_{w}^{\omega,\ast} \}$ into \textbf{RP2}, and denote the minimized actual operating cost as
\begin{equation}
 \hat{f}(\alpha_{s},\alpha_{w},\alpha_{e},\omega)= \min_{\boldsymbol{Q}^{\omega},\boldsymbol{x}_{i}^{\omega}} 
 [ \hat{C}_{o}(\boldsymbol{Q}^{\omega},\boldsymbol{e}_{s}^{\omega,\ast},\boldsymbol{e}_{w}^{\omega,\ast}) 
 + \sum_{i \in \mathcal{N}} C_{i} (\boldsymbol{x}_{i}^{\omega}) 
 ]. \label{actualsocialcost}
\end{equation}

Based on the minimized actual operating cost \eqref{actualsocialcost} in period-2, we write the worst-case overall cost minimizing problem in period-1 as
\begin{align*}
& \leftline{\textbf{RP1: Robust optimization for overall cost minimization}}
\end{align*}
\begin{equation*}
 \begin{aligned}
 &\underset{\alpha_{s},\alpha_{w},\alpha_{e}}{\min}\;
 && C_{I}(\alpha_{s},\alpha_{w},\alpha_{e}) 
 +  D \cdot \mathbb{E}_{\omega \in \Omega}\left[ \hat{f}(\alpha_{s},\alpha_{w},\alpha_{e},\omega) \right] \\
 &\text{subject to}
 && \text{Constraints \eqref{investconstraint1} and \eqref{investconstraint2}},
 \end{aligned}
\end{equation*}
which solves the optimal capacity investment under the worst-case prediction of renewable energy scenarios. Note that \textbf{RP1} shares the same structure as Problem \textbf{P1}, and thus can be solved by the same methodology presented in Section VII.

\section{Simulation Results}
We assume that the investment horizon includes $D=3650$ days (10 years). We use the load curve in \cite{loadcurve} as the preferred power consumption, and set $\beta_{o}=0.005$ and $\beta_{i}=0.5$ for $i \in \mathcal{N}$. The investment costs of solar energy, wind energy and energy storage are set as $c_s=12,480$ HKD per kW, $c_w=7,800$ HKD per kW and $c_e=1,950$ HKD per kWh. We obtain renewable power scenarios as discussed in Section II, with details described in the online technical report \cite{report}.

\subsection{Optimal investment}
\begin{figure}[tbhp]
	\centering
	\includegraphics[width=6.5cm]{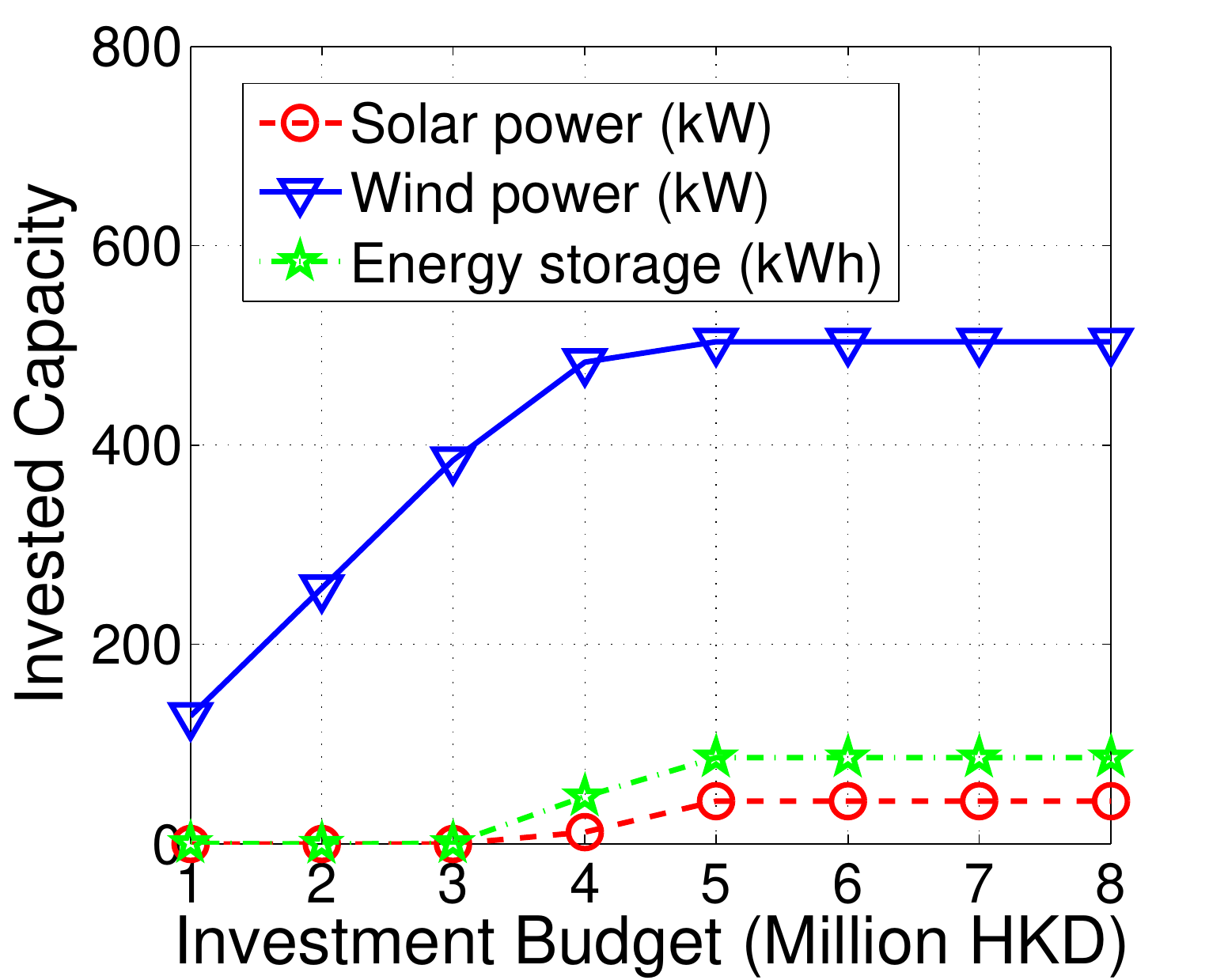}
	\caption{\label{fig-invest-TC}Capacity investment at TC}
\end{figure}

\begin{figure}[tbhp]
	\centering
	\includegraphics[width=6.5cm]{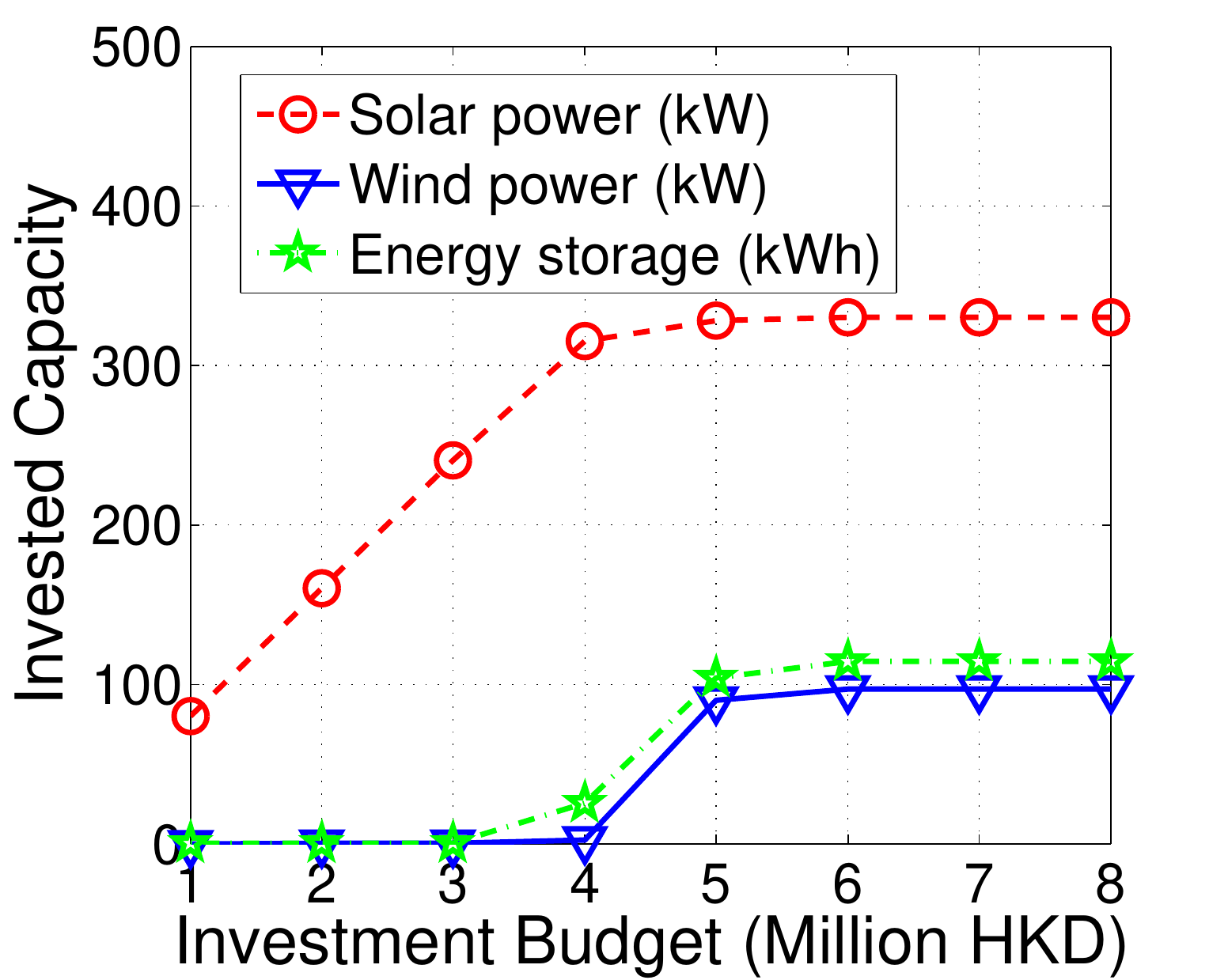}
	\caption{\label{fig-invest-SKG}Capacity investment at SKG}  
\end{figure}

First, we study the optimal investment strategies of solar power, wind power, and energy storage in TC and SKG under different budget constraints, as shown in Fig. \ref{fig-invest-TC} and Fig. \ref{fig-invest-SKG}, respectively. When the budget is tight (\emph{e.g.}, below 3 million HKD), the investment priority at location TC is wind power as it's more economical. While the budget increases, investments in solar power and energy storage capacities also increase, but wind power still dominates the energy portfolio. This is because the wind power output is higher than solar power at location TC, and the demand response scheme as well as energy storage can help better utilize the wind power. The optimal investment expense is 4.6 million HKD at TC, and even given more budget (\emph{e.g.}, greater than 5 million HKD), the optimal capacity investment remains the same. From Fig. \ref{fig-invest-SKG}, we see that the optimal investment strategy at location SKG is different, with the main difference that more investment is put into solar power generation. This is because the wind power output at SKG is not as adequate as that at TC and solar power fits the daily demand better at SKG. As a result, wind power generation becomes less competitive at SKG. When the budget is less than 4 million HKD, all the investment goes to solar power, and when the budget further increases, the investments in wind power and energy storage increase. The optimal investment expense is 5.1 million HKD at SKG.

\subsection{Benefit of demand response} 
\begin{figure}[tbhp]
	\centering
	\vspace{-2mm}
	\includegraphics[width=6.0cm]{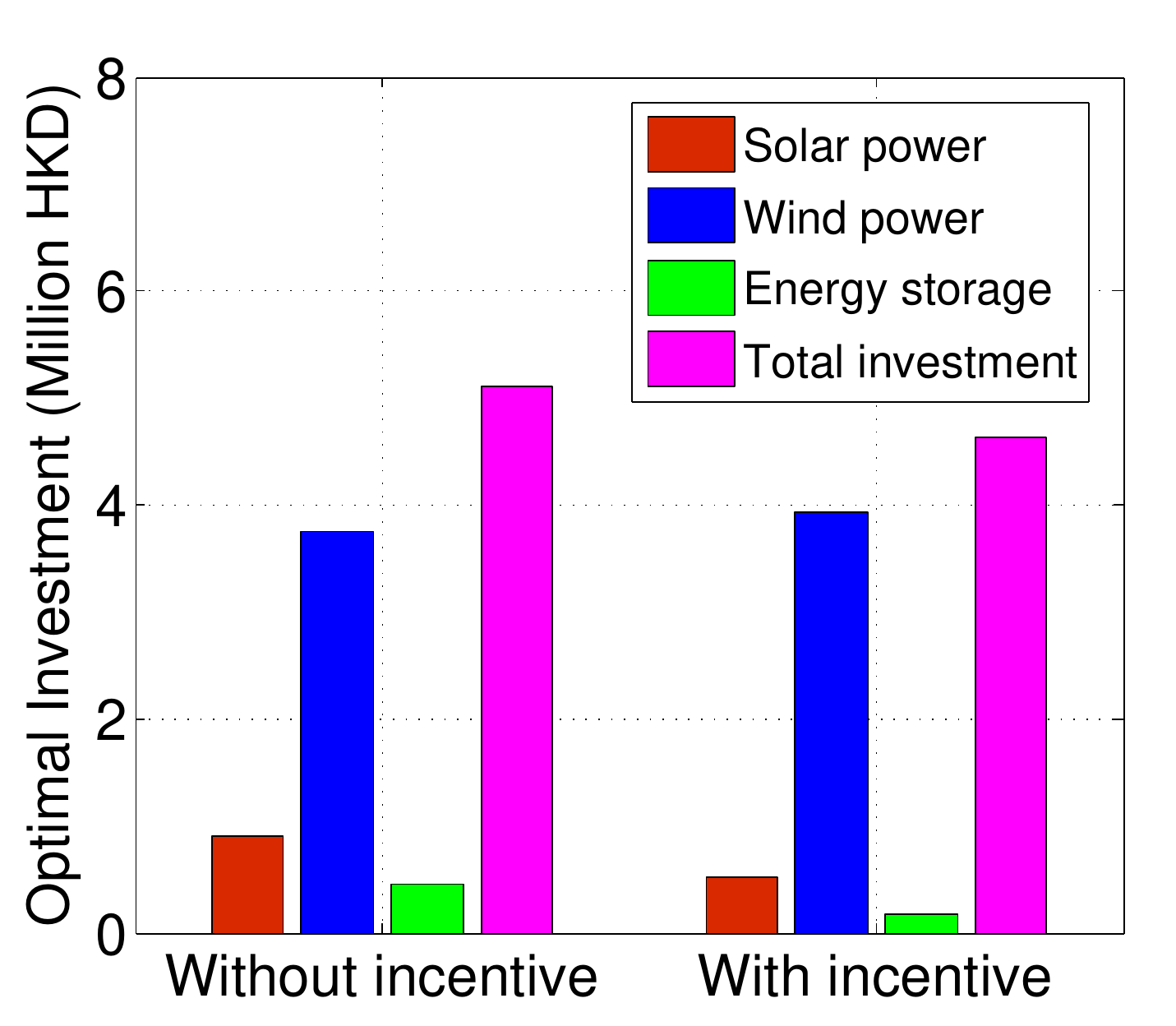}
	\caption{\label{fig-incentive-TC}Energy portfolio investment at TC}
	\vspace{-4mm}
\end{figure}

\begin{figure}[tbhp]
	\centering
	\includegraphics[width=6.0cm]{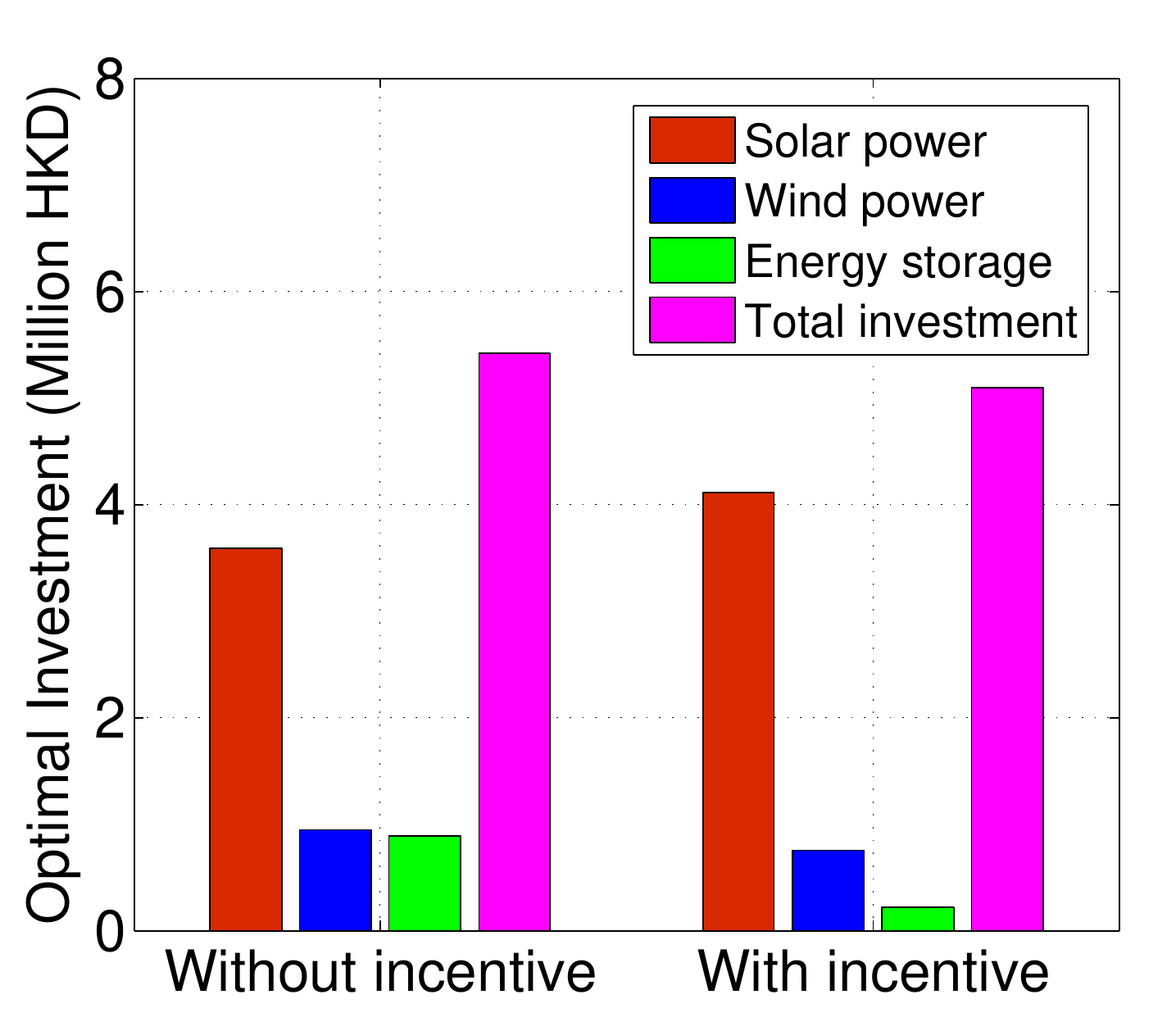}
	\caption{\label{fig-incentive-SKG}Energy portfolio investment at SKG}
\end{figure}

We study the benefit of adopting demand response schemes, and depict the results in Fig. \ref{fig-incentive-TC} and Fig. \ref{fig-incentive-SKG} for locations TC and SKG, respectively. The investment budget in period-1 is set as 6 million HKD. At location TC, \emph{without} incentives (in which case we keep the prices low and the same for 24 hours), a user $i$ will choose the power consumption according to the preferred power consumption $\boldsymbol{y}_i$. In that case, it is optimal for the operator to use 5.1 million HKD budget for investment in period-1. \emph{With} price incentives, the operator sets time dependent day-ahead prices so that to steer the users' power consumptions to the socially optimal values. Comparing with the case without incentives, optimal demand response under incentives enables the operator to invest more wind power, which has a lower investment cost than solar power. Energy storage investment also decreases with incentives, because the elastic demand can be shifted proactively, which reduces the dependence on the energy storage. The total optimal investment expenditure reduces by 9.4\% to 4.6 million HKD, which implies that the demand response may significantly reduce the system cost and avoid over-investment.

At location SKG, Fig. \ref{fig-incentive-SKG} shows that it is optimal to use 5.4 million HKD budget for investment in period-1 \emph{without} incentives. However, \emph{with} price incentives, optimal demand response under incentives enables the operator to invest more solar power but less energy storage, because the solar power fits the demand better compared with the wind power at SKG. The total optimal investment expenditure reduces by 6.1\% to 5.1 million HKD, demonstrating the benefit of demand response.

\subsection{Optimal power scheduling and pricing}
\begin{figure}[tbhp]
	\centering
	\vspace{-2mm}
	\includegraphics[width=6.0cm]{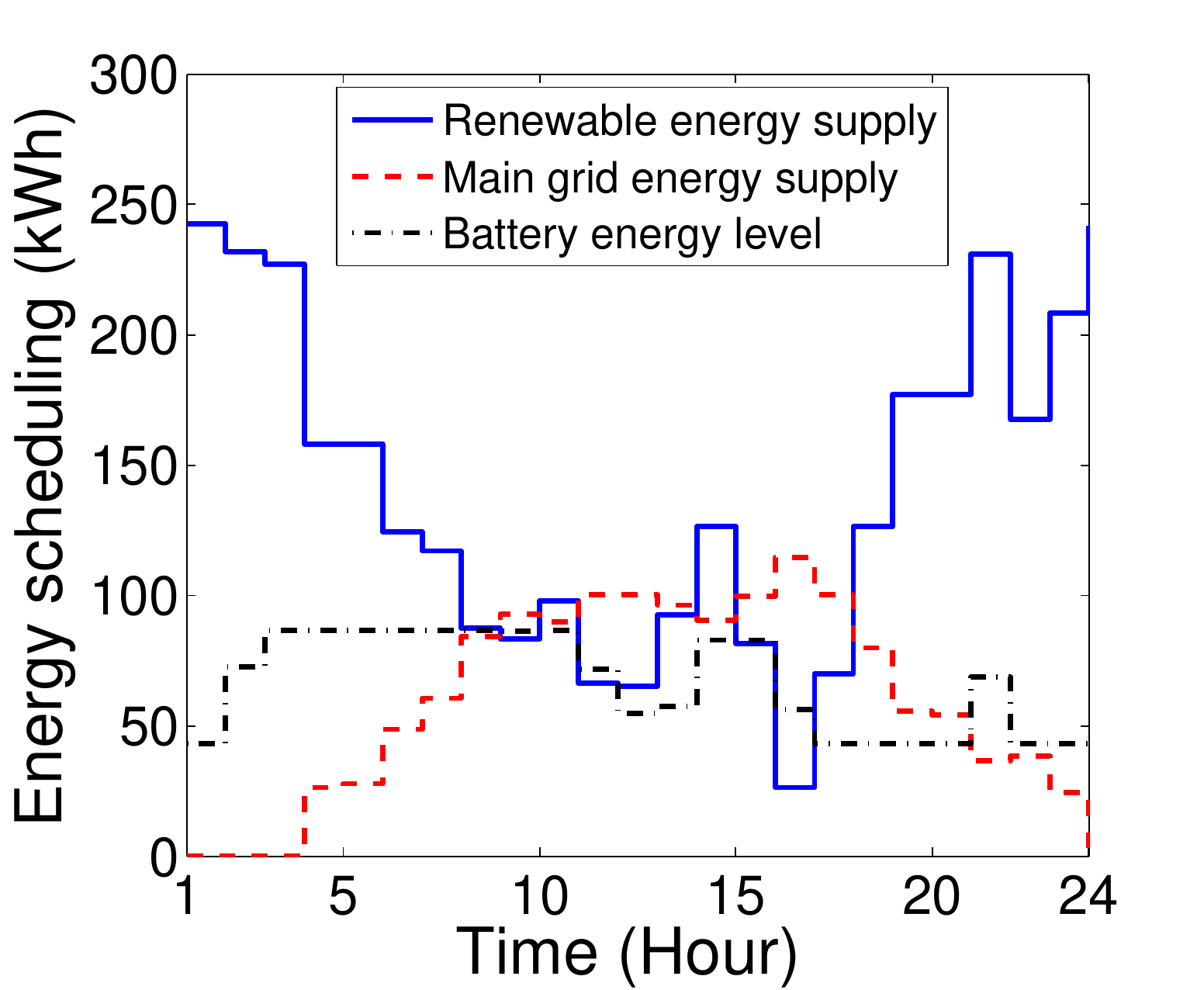}
	\caption{\label{fig-scheduling-TC}Energy supply and storage at TC}
	\vspace{-4mm}
\end{figure}

\begin{figure}[tbhp]
	\centering
	\includegraphics[width=6.0cm]{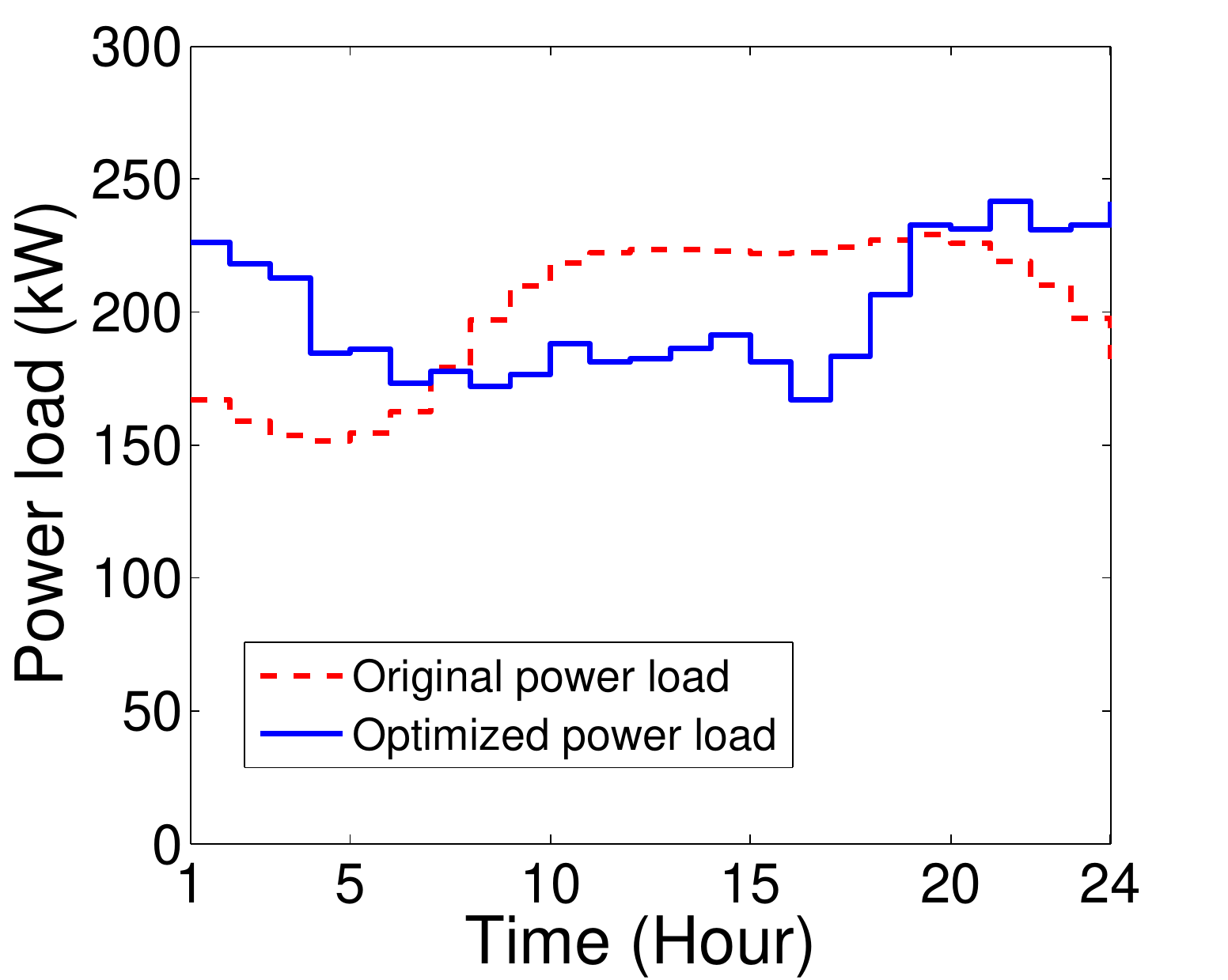}
	\caption{\label{fig-powerload-TC}Power load at TC}
\end{figure}

Next we focus on the numerical studies of the power scheduling in period-2. First, we plot the energy supply and power load at location TC of a typical day in Fig. \ref{fig-scheduling-TC} and Fig. \ref{fig-powerload-TC}, respectively. Note that we treat the aggregate supply from both solar power and wind power as renewable power in period-2. Fig. \ref{fig-scheduling-TC} shows that the renewable energy generation provides high energy supply, especially at night time during hour 7PM-5AM. Renewable energy generation drops drastically during hour 9AM-5PM, because location TC is mainly supplied by wind power and wind power produces less at day time. During this time period, the microgrid operator needs to compensate the loss of renewable energy generation through discharging the energy storage and purchasing energy from the main grid, as we can see that the energy storage level decreases. From Fig. \ref{fig-powerload-TC}, we see that the users respond to the day-ahead prices by optimizing the power load (demand), so that to shift the original peak power load during hour 9AM-5PM to other time slots when there is more renewable energy available.

\begin{figure}[tbhp]
	\centering
	\vspace{-2mm}
	\includegraphics[width=6.0cm]{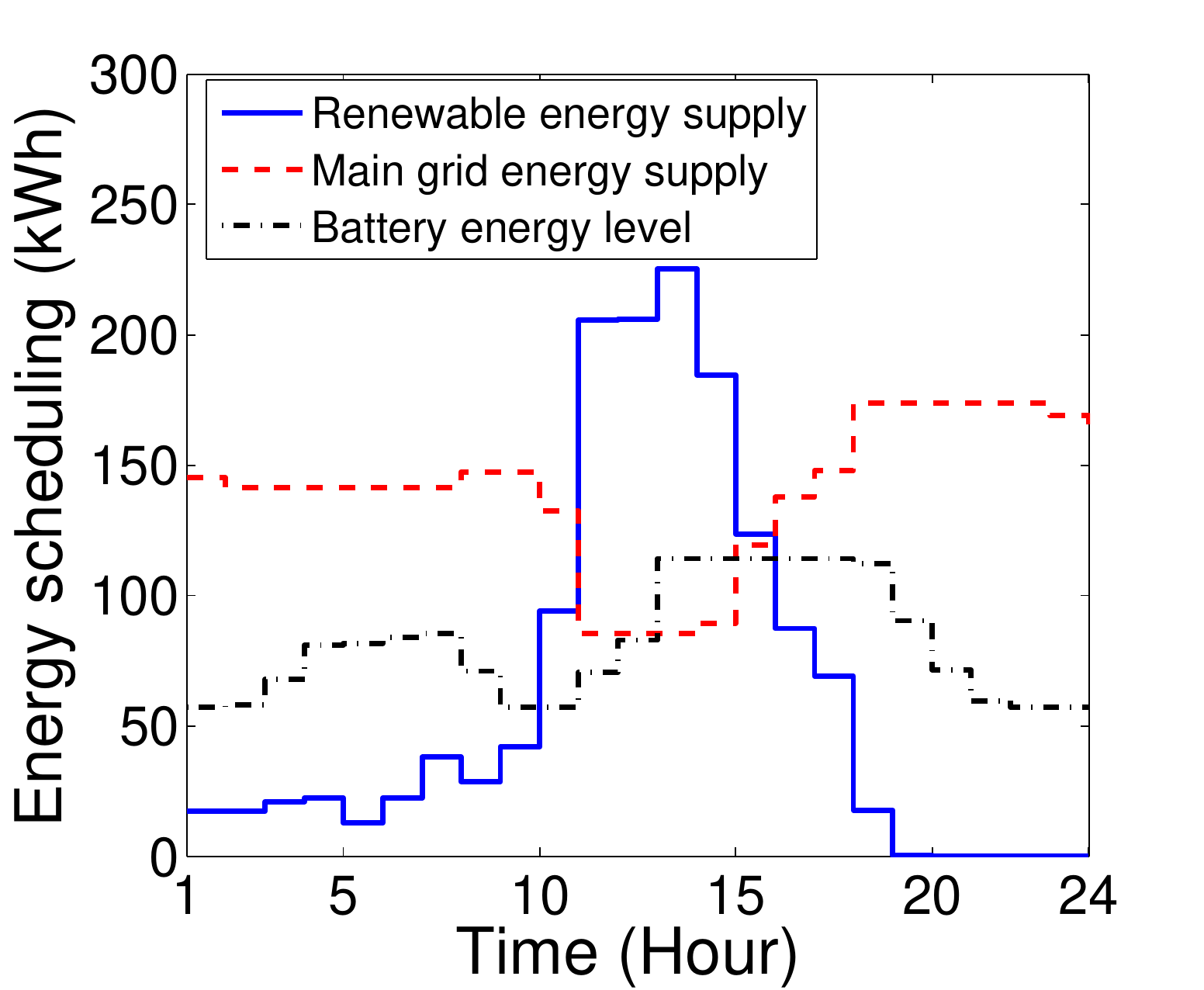}
	\caption{\label{fig-scheduling-SKG}Energy supply and storage at SKG}
	\vspace{-4mm}
\end{figure}

\begin{figure}[tbhp]
	\centering
	\includegraphics[width=6.0cm]{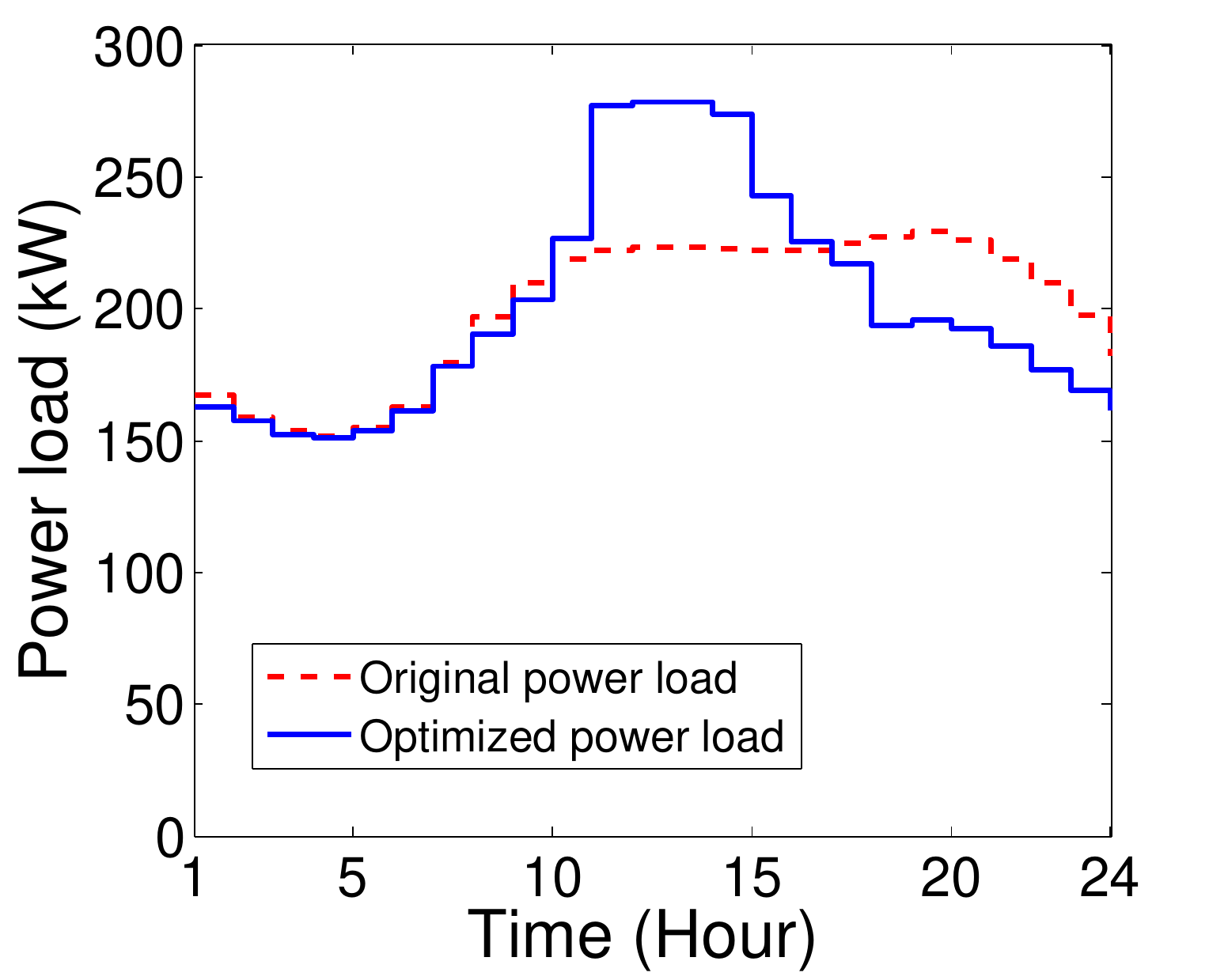}
	\caption{\label{fig-powerload-SKG}Power load at SKG}
\vspace{-3mm}
\end{figure}

We also study the energy supply and power load of a typical day at location SKG, shown in Fig. \ref{fig-scheduling-SKG} and Fig. \ref{fig-powerload-SKG}, respectively. Fig. \ref{fig-scheduling-SKG} shows that the renewable energy generation provides high supply during day-time, especially during hour 10AM-4PM. During other time slots, the microgrid operator needs to make sure all the demand can be satisfied, and thus purchases energy from the main grid. The microgrid operator also charges energy storage during high renewable supply periods (11AM-1PM), and discharges the energy storage afterwards as we can see that the energy storage level decreases after hour 7PM. Since location SKG has a large portion of solar power, and there is a power supply peak during day-time, we see from Fig. \ref{fig-powerload-SKG} that the original load during evening-time after hour 6PM is shifted to day-time when there is more renewable energy available. 

\begin{figure}[tbhp]
	\centering
	\vspace{-2mm}
	\includegraphics[width=6.0cm]{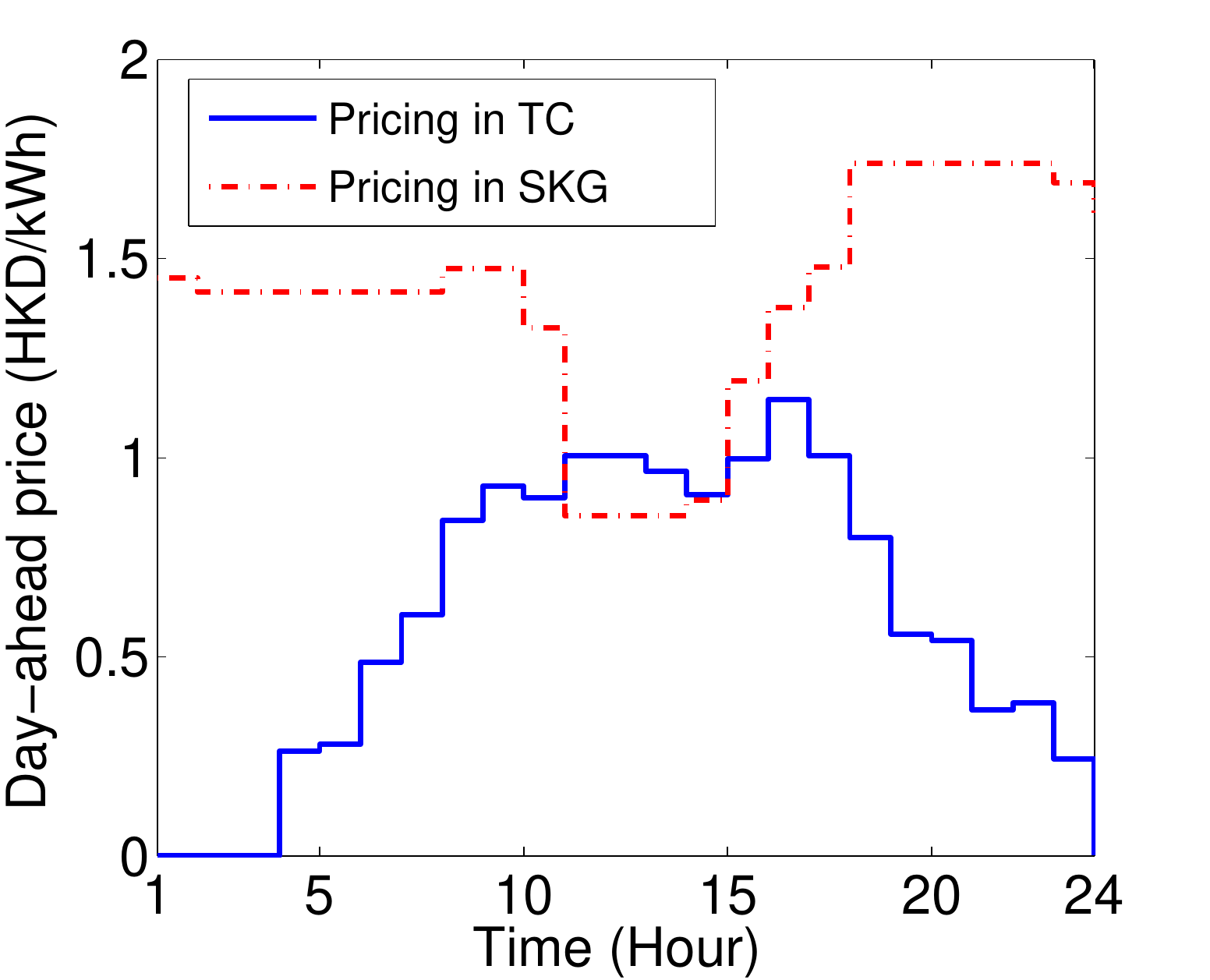}
	\caption{\label{fig-price} Day-ahead prices at TC and SKG}
    \vspace{-4mm}
\end{figure}

Fig. \ref{fig-price} shows the internal day-ahead prices at TC and SKG. The prices indicate the marginal operating cost of the microgrid. We see that the microgrid operator does not charge the users when the demand can be satisfied by local renewable energy generation (\emph{e.g.}, hour 1AM-3AM at TC). At location TC, the operator charges high prices during day time, but low prices at night time. The reason is that the renewable portfolio consists of more wind power at TC, which has higher power output at night than that during day time. The operator needs to import more main grid power during hour 9AM-6PM, which incurs operating cost. At location SKG, the price trend is opposite to the price at TC. Because more solar power is invested at SKG, which produces more renewable power during day time. However, during night time, when there is no solar power supply, the operator relies on main grid power, and thus charges relatively high prices at night. The day-ahead prices set by the operator provide efficiently incentivize users to shift their flexible loads. In Fig. \ref{fig-powerload-TC} and Fig. \ref{fig-powerload-SKG}, we see that users' flexible loads deviate from the original patterns and are shifted from high-price time slots to other low-price time slots at both TC and SKG.

\subsection{Impact of prediction errors}
\begin{figure}[tbhp]
	\centering
	\vspace{-2mm}
	\includegraphics[width=6.0cm]{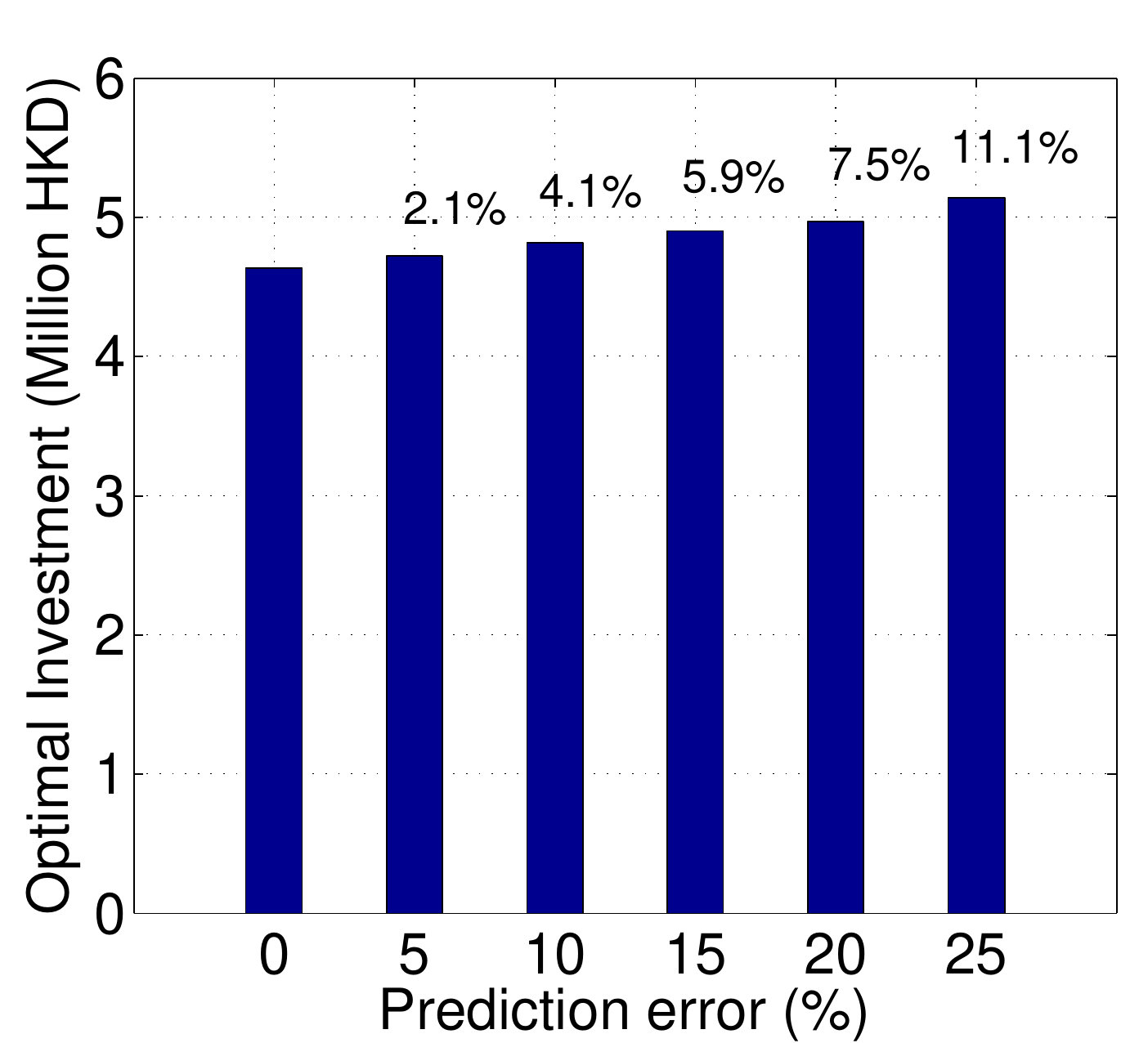}
	\caption{\label{fig-error} Impact of prediction errors}
	\vspace{-6mm}
\end{figure}

We take location TC as an example to study the impact of prediction errors on the optimal investment, as shown in Fig. \ref{fig-error}. When there is zero prediction error of renewable generations, the optimal investment is 4.6 million HKD. With the increase of prediction errors, the optimal investment expense increases, as the operator over-invests to hedge the risk of renewable generation shortage caused by prediction errors. For example, when the error is 10\%, the operator invests 4.1\% over the 4.6 million HKD benchmark. This shows the importance of accurate predication for achieving an optimal investment decisions. 

\subsection{Comparison with existing studies}
We add new simulations to compare our proposed method with existing studies in the simulation. Specifically, we consider four benchmark methods, which are motivated by \cite{invest3}, \cite{invest4}, \cite{invest5} and \cite{invest6}, respectively. We first compute the optimal capacity investment for all the benchmark methods, and then calculate the corresponding investment cost and the actual operational cost using one-year renewable energy data. The simulation results (including the optimal capacity investment, investment cost, operational cost, and overall cost) are summarized in Table \ref{table}. 

\begin{table*}
	\centering
	\caption{Performance comparison}
	\label{table}
	\begin{tabular}{ |p{2.0cm}||p{1.2cm}|p{1.5cm}|p{1.5cm}|p{1.8cm}|p{1.8cm}|p{2.0cm}|p{1.8cm}|  }
		\hline
		Methods & Demand response & Solar power (kW) & Wind power (kW) & Energy storage (kWh) & Investment cost (Million HKD) & Operational cost (Million HKD) & Overall cost (Million HKD) \\
		\hline
		Benchmark 1 \cite{invest3}  & N & 304.4 & N & 473.1 & 4.7 & 14.2 & 18.9 \\
		\hline
		Benchmark 2 \cite{invest4} & N & N & 536.9 & 298.4 & 4.8 & 2.8 & 7.6 \\
		\hline
		Benchmark 3 \cite{invest5}  & N & 77.8 & 494.6 & N & 4.8 & 3.6 & 8.4 \\
		\hline
		Benchmark 4 \cite{invest6}  & N & 71.8 & 481.4 & 233.1 & 5.1 & 2.5 & 7.6 \\
		\hline
		Our method   & Y & 42.4 & 503.5 & 86.5 & 4.6 & 2.0 & 6.6 \\
		\hline
	\end{tabular}
\end{table*}

Benchmark 1 method focuses on the solar-storage investment, and benchmark 2 method focuses on the wind-storage investment. We see that benchmark 2 achieves a much lower overall cost than benchmark 1 (7.6 vs. 18.9 Million HKD), because wind power is more abundant than solar power, and the investment cost of wind power per unit capacity is also lower than that of the solar power. Benchmark 3 and benchmark 4 show the mixed renewable energy investment with and without energy storage, respectively. We see that benchmark 4 achieves a lower overall cost than benchmark 3 (7.6 vs. 8.4 Million HKD), because the energy storage can help deal with the stochastic nature of renewable energy by proper charging and discharging. Our proposed method considers a comprehensive configuration of microgrids (solar power, wind power, and energy storage), and also optimizes demand response. Hence our proposed method has the minimum investment cost and operational cost, and achieves the minimum overall cost compared with all other benchmark methods. The simulation results in Table \ref{table} demonstrate the advantage of our proposed method over the existing literature.

\section{Conclusion}
We proposed a theoretical framework to study the joint investment and operation problem in the microgrid. The two-period stochastic program models the renewable energy uncertainty, and captures the coupled nature of investment and operation. With realistic meteorological data, our model provides the optimal investment decisions on renewable energy and energy storage capacities, and the optimal demand response scheme (pricing and power scheduling) in a microgrid. The simulation studies demonstrate the economic benefit of demand response and the impact of prediction accuracy.

\end{document}